\documentclass[sigconf, 9pt]{acmart}
\newcommand{\cmark}{\ding{51}}%
\newcommand{\xmark}{\ding{55}}%

\newcommand{\sys}{CacheSolidarity\xspace}

\newenvironment{tightitemize}%
	 {\begin{list}{$\bullet$}{%
 		\setlength{\leftmargin}{10pt}
        \setlength{\itemsep}{0pt}%
        \setlength{\parsep}{0pt}%
        \setlength{\topsep}{0pt}%
        \setlength{\parskip}{0pt}%
        }%
 }%
{\end{list}}

\newcommand{\citeaut}[1]{{\citeauthor{#1}~\cite{#1}}}

\newcommand{\compactpara}[1]{\smallskip\noindent\textbf{#1}}

\usepackage{enumitem}
\usepackage{graphicx}
\usepackage{caption}
\usepackage{subcaption}
\usepackage{comment}
\usepackage{pifont}% http://ctan.org/pkg/pifont
\usepackage{algorithm}
\usepackage{algpseudocode}
\usepackage{amsmath}
\usepackage{xspace}
\usepackage[switch]{lineno}
\usepackage{tcolorbox}
\usepackage{amsthm}
\usepackage{hyperref}

\usepackage[normalem]{ulem}
\usepackage{xcolor}
\usepackage{tabularx}
\usepackage{seqsplit}

\usepackage{listings}

\graphicspath{ {./figures/} }

\title{\sys: Preventing Prefix Caching Side Channels in Multi-Tenant LLM Serving Systems}

\author{Panagiotis Georgios Pennas}
\affiliation{%
  \institution{IMDEA Software Institute}
  \institution{Universidad Politécnica de Madrid}
  \city{Madrid}
  \country{Spain}}
\email{panagiotis.pennas@imdea.org}

\author{Konstantinos Papaioannou}
\affiliation{%
  \institution{IMDEA Software Institute}
  \institution{Universidad Politécnica de Madrid}
  \city{Madrid}
  \country{Spain}}
\email{konstantinos.papaioannou@imdea.org}

\author{Marco Guarnieri}
\affiliation{%
  \institution{IMDEA Software Institute}
  \city{Madrid}
  \country{Spain}}
\email{marco.guarnieri@imdea.org}

\author{Thaleia Dimitra Doudali}
\affiliation{%
  \institution{IMDEA Software Institute}
  \city{Madrid}
  \country{Spain}}
\email{thaleia.doudali@imdea.org}

\copyrightyear{2026}
\acmYear{2026}
\setcopyright{acmlicensed}
\acmConference[Middleware '26]{27th ACM/IFIP International Middleware Conference}{December 2026}{Tarragona, Spain}
\acmBooktitle{27th ACM/IFIP International Middleware Conference (Middleware '26), December 2026, Tarragona, Spain}
\acmDOI{10.1145/XXXXXXX.XXXXXXX}
\acmISBN{979-8-XXXX-XXXX-X/26/12}

\copyrightyear{2026}
\acmYear{2026}
\setcopyright{cc}
\setcctype{by}
\acmConference[Middleware '26]{27th International Middleware Conference}{December 14--18, 2026}{Tarragona, Spain}
\acmBooktitle{27th International Middleware Conference (Middleware '26), December 14--18, 2026, Tarragona, Spain}
\acmDOI{10.1145/3801927.3858638}
\acmISBN{979-8-4007-2621-7/2026/12}

\begin{document}

% Abstract
\begin{abstract}
Large Language Models (LLMs) rely on optimizations like Automatic Prefix Caching (APC) to accelerate inference.
APC works by reusing previously computed states for the beginning part of a request (prefix), when another request starts with the same text.
While APC improves throughput, it introduces timing side channels: cache hits are faster than misses, creating observable latency differences. 
In multi-tenant systems, attackers can exploit these differences to infer sensitive information, e.g., by incrementally reconstructing another user's request by observing hit/miss patterns.
Current defenses take a sledgehammer approach: they disable APC and cache sharing, isolating users, and sacrificing efficiency for regular users. 
%to eliminate potential timing leaks. 
%
%While this guarantees security, it erases APC's performance benefits and penalizes regular users, leading to inefficiency and inability to scale. 
%
This paper presents \sys, a system that secures multi-tenant LLM serving systems against APC side channels without sacrificing performance and efficiency.
%
%\sys monitors cache access patterns across users, flags suspicious reuse of cached shared prefixes, and selectively isolates later paths of flagged prefixes using minimal extensions to KV cache entries, thereby isolating only those requests that might be involved in attacks.
\sys monitors cache reuse across users, flags suspicious sharing, and selectively isolates prefixes, restricting their reuse only when necessary.
Evaluation shows that \sys enables up to 70\% higher cache reuse and 30\% lower inference latency compared to existing user-isolation defenses. %that isolate users. 
\sys's lightweight design demonstrates %how security in LLM serving does not have to come at the cost of unnecessarily reduced performance or unbearable overheads.
how LLM serving can be secured without sacrificing performance or incurring unbearable overheads.
\end{abstract}




\begin{comment}
% Title
% \settopmatter{printfolios=true}
    \settopmatter{printfolios=true,printacmref=false}
    \renewcommand\footnotetextcopyrightpermission[1]{} % removes footnote with conference information in first column
    \maketitle
    \pagestyle{plain}
\end{comment}

% Title
\maketitle

% Sections
\section{Introduction}
\label{sec:intro}

Large Language Models (LLMs) now power applications such as conversational assistants, code generation, and enterprise analytics~\cite{brown2020language}. These services operate at massive scale under strict latency and throughput requirements, making inference serving a critical systems challenge~\cite{dist-serve}. Modern serving systems employ system-level optimizations such as cache management~\cite{kwon2023pagedattention,infini-gen,llumnix} and scheduling~\cite{sarathi,sarathi-serve,deepspeed-fastgen} to accelerate LLM inference. Among them, prefix sharing~\cite{juravsky2024hydragen,chunk-attention, bang2023gptcache}, or {\it Automatic Prefix Caching (APC)}, is widely used by state-of-the-art frameworks and commercial APIs, such as  OpenAI~\cite{openai2024promptcaching}, DeepSeek~\cite{deepseek2024contextcaching}, Google Gemini~\cite{deepmind2025gemini}, MoonShot Kimi~\cite{qin2025mooncake}, vLLM~\cite{kwon2023pagedattention}, and SGLang~\cite{zheng2024sglang}. 
APC accelerates inference by caching and reusing previously computed model states for the beginning part of a request (the request's \textit{prefix}) whenever another request starts with the same text. This avoids redundant computation and significantly reduces latency for long prompts and multi-turn conversations~\cite{juravsky2024hydragen,chunk-attention}.

While APC improves performance, it introduces security risks similar to other caching optimizations~\cite{flushreload,dnscache,pagecache,browsercache}. APC creates observable inference latency differences between cache hits and misses, resulting in a \textit{timing side channel} that can reveal sensitive information. In multi-tenant deployments where prefix caches are shared across users, attackers can exploit these differences to infer private data. For example, by sending crafted requests and measuring time-to-first-token (TTFT), an attacker can determine whether their input matches cached prefixes from another user's request. Repeated probing enables prompt-stealing attacks that reconstruct sensitive content from other users~\cite{early-bird,pmlr-v267-gu25b}.

To prevent these attacks, prior work proposes three main defenses against APC timing leaks: (i) full user-level isolation, which disables prefix sharing entirely~\cite{cache-partitioning,input-snatch,pmlr-v267-gu25b}; (ii) timing obfuscation, which injects noise to mask latency differences~\cite{remote-attacks}; and (iii) selective isolation of secret-dependent prompts based on LLM-aided semantic analysis~\cite{chu2026selectivekvcachesharingmitigate}. 
While effective in theory, all these defenses suffer from a critical drawback: they introduce significant overhead, \emph{often unnecessarily}, e.g., for prompts that are not involved in attacks or for situations where the timing side channel is not exploitable in practice. Thus, these solutions erase APC’s performance benefits and penalize regular, benign users; thereby limiting their applicability in realistic, high-performance multi-tenant environments.

To address the shortcomings of current defenses, we introduce {\bf \sys}, a system-level solution that secures multi-tenant LLM serving against timing side channels caused by APC, while preserving its performance benefits. \sys is based on the {\bf key insight} that isolating entire users (as done in~\cite{cache-partitioning, input-snatch, pmlr-v267-gu25b}) is not necessary to prevent attacks; rather {\it it is sufficient to protect the shared prefixes that might lead to attacks.} Building on this insight, \sys{} (1) continuously monitors prefix reuse across users, (2) flags prefixes that are suspiciously reused by multiple users, and (3) prevents further reuse of flagged prefixes in follow-up requests. This approach embodies the principle of cooperative efficiency: benign users benefit from shared caching through \textit{solidarity}, where common prefixes remain accessible to all. When suspicious behavior threatens this solidarity, \sys intervenes by isolating only the flagged prefixes instead of penalizing all users.

%The system design of \sys{} consists of lightweight components that work together to secure prefix caching without sacrificing efficiency. First, \sys{} extends the prefix cache with minimal metadata for each entry to track ownership and flag suspicious reuse. Second, the {\it Detector} monitors cache hits and isolates flagged prefixes when multiple users attempt to share them. Finally, \sys{} introduces a system-level optimization that activates prefix isolation only when timing side channels are actually exploitable. This is handled by the {\it Activator}, which continuously evaluates whether latency differences between cache hits and misses are distinguishable, based on factors like the hardware platform, LLM model size, system load, and request length. These parameters directly influence the strength and exploitability of the side channel, as demonstrated in our analysis in \S\ref{sec:motivation}. Overall \sys's system components interact seamlessly: the cache extension provides metadata, the Detector enforces selective isolation, and the Activator optimizes the security-performance trade-off.\looseness=-1 
\sys{} consists of lightweight components that work together to secure prefix caching without sacrificing efficiency. First, \sys{} extends the prefix cache with minimal metadata to track ownership and flag suspicious reuse. Second, the {\it Detector} monitors cache hits and isolates flagged prefixes shared across users. Finally, the {\it Activator} enables isolation only when timing side channels are exploitable, based on the distinguishability of cache-hit and cache-miss latencies. This depends on factors such as hardware, LLM model size, system load, and request length, as shown in Section~\ref{sec:motivation}. Together, \sys's system components interact seamlessly: the cache extension provides metadata, the Detector enforces selective isolation and the Activator optimizes the security-performance trade-off.

%\sys{} is implemented on top of the open-source, state-of-the-art LLM serving system vLLM~\cite{vllm}. We conduct an extensive evaluation across multi-tenant workloads with varying levels of intra- and inter-user prefix sharing and 10 LLM models spanning different families and sizes (from 0.5B up to 70B parameters). Our experimental results show that \sys{} achieves up to 70\% higher cache reuse and 30\% lower inference latency compared to defenses that enforce user-level isolation. In addition, \sys{} introduces negligible time and memory overheads during runtime, demonstrating that security can be achieved without sacrificing performance when implemented in a lightweight and principled way. To validate the security of \sys{}, we performed a detailed security analysis that precisely characterizes \sys{}'s security guarantees, which we complement with an empirical validation showing that \sys{} indeed closes the timing side-channel introduced by APC across requests from different users.\looseness=-1
\sys{} is implemented on top of the state-of-the-art LLM serving system vLLM~\cite{vllm}. Extensive evaluation across workloads and LLM model families and sizes shows that \sys{} achieves up to 70\% higher cache reuse and 30\% lower inference latency than user-level isolation defenses, with negligible runtime and memory overhead. We further characterize \sys{}'s security guarantees through a detailed analysis, complemented by empirical validation demonstrating its effectiveness against APC timing side channels across users.

\begin{figure}[t]
    \centering
    \includegraphics[width=\columnwidth]{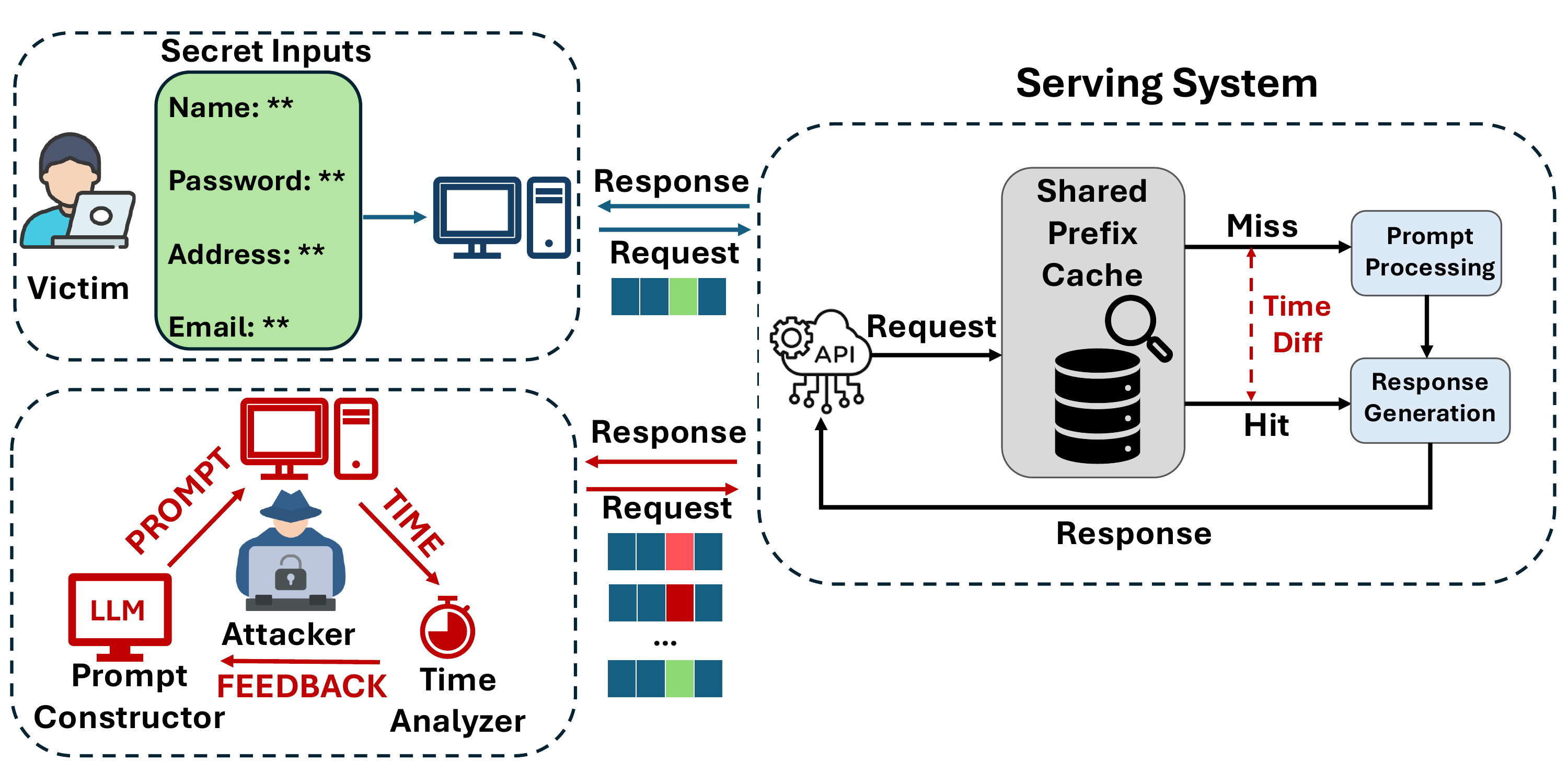}
    \vspace{-0.2in}
    \caption{Timing side-channel leakage in prefix-sharing LLM inference. The attacker sends crafted prompts and measures time-to-first-token (TTFT) to detect cache hits or misses caused by Automatic Prefix Caching (APC) and steal the sensitive information in the victim's prompt.}
    \label{fig:attack}
    \vspace{-0.3in}
\end{figure}

% \vspace{-0.03in}
\noindent The specific {\bf paper contributions} are:
\begin{tightitemize}
    \item An analysis of APC timing side channels in multi-tenant LLM serving systems, identifying the workload and system conditions affecting their exploitability (Section~\ref{sec:motivation}).
    \item The design of {\bf \sys}, a lightweight system that prevents timing attacks by selectively isolating suspicious prefixes rather than entire users, preserving caching benefits (Section~\ref{sec:system}). The code is open-sourced.\footnote{{\url{https://github.com/muse-research-lab/CacheSolidarity}}}
    \item A security analysis of \sys{}'s guarantees and limitations, complemented by empirical validation against prompt-stealing attacks (Section~\ref{sec:system-sec-analysis}).
    \item An extensive evaluation of the performance of \sys across diverse workloads and state-of-the-art LLM models, alongside a sensitivity analysis (Section~\ref{sec:evaluation}).
\end{tightitemize}
% \vspace{-0.03in}

\section{Background and Motivation}
\label{sec:motivation}

In this section, we provide background and motivation for \sys{}.
We start by introducing background on LLM serving systems and, in particular, on the APC optimization (\ref{sec:motivation:llm-serving-systems}).
We then characterize the parameters that influence the exploitability of timing side channels due to APC (Section~\ref{sec:motivation:leaks}).
Then, we illustrate how attackers can exploit APC through timing leaks to learn sensitive prompts (\ref{sec:motivation:attacks}).
Next, we overview current defenses for APC timing leaks and discuss their limitations  (\ref{sec:motivation:defenses}).
%
%We conclude by summarizing our motivational observations in (\ref{sec:motivation:summary}).

%%%%%%%%%%%%%%%%%%%%%%%%%%%%%%%%%%%%%%%%%%%%%%%%%%%%%%%%%%%%%%%%%%%%%%%%%%%%%%%%%%%%%%%%%%%%%%%%%%%%%%%%%%%%%%%%
\subsection{Caching in LLM Serving Systems}\label{sec:motivation:llm-serving-systems}

Large Language Models (LLMs), such as GPT~\cite{brown2020language} and LlaMA~\cite{touvron2023llama}, are the backbone of modern AI services.
Integrating them in modern IT systems requires serving systems that deliver inference with low latency and high throughput under multi-user workloads. 
Inference happens in two stages: \textit{prefill} and \textit{decode}. 
During prefill, the entire input prompt (also called a \emph{request}) is processed in a single forward pass, generating key-value (KV) tensors for each token. 
In the decode phase, tokens are produced autoregressively; for each new token, the model computes a new key-value tensor using all previously generated tensors. 
Without caching, every decode step would recompute all KV tensors for both prefill and prior decode tokens, incurring quadratic cost. 

To avoid this, current LLM serving systems employ a KV cache that stores tensors generated during the prefill and previous decoding steps for reuse. To further accelerate inference, several commercial frameworks such as OpenAI~\cite{openai2024promptcaching}, DeepSeek~\cite{deepseek2024contextcaching}, Google Gemini~\cite{deepmind2025gemini}, MoonShot Kimi~\cite{qin2025mooncake}, vLLM~\cite{kwon2023pagedattention}, and SGLang~\cite{zheng2024sglang} implement \emph{Automatic Prefix Caching (APC)}. APC reduces redundant computation by reusing cached KV tensors whenever a new request shares a prefix with an earlier one. This optimization is particularly effective for scenarios such as long-document queries or multi-round conversations~\cite{vllm_prefix_caching}, where repeated processing of the same prefix would otherwise incur significant overhead.

When serving systems apply APC, a prompt can be viewed as {\it a sequence of prefixes}, each mapped to a cache entry in the KV/prefix cache. Requests may experience partial cache hits, starting from the beginning of the prompt and reusing some cached prefixes while recomputing others. The leftmost example in Figure~\ref{fig:example} illustrates this process, where each node corresponds to a cache entry for that particular prefix/part of the prompt sentence.

\begin{figure}[t]
    \centering
    
    \begin{subfigure}[t]{1.0\columnwidth}
        \centering
        \includegraphics[width=0.5\columnwidth]{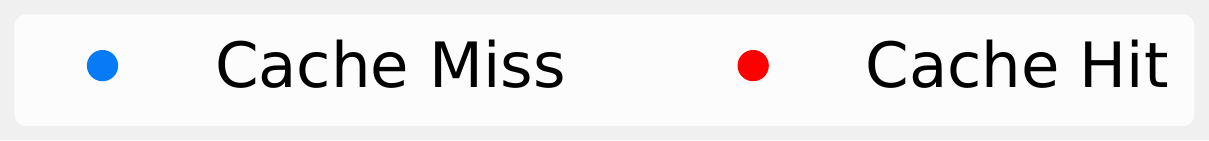}
    \end{subfigure}
    % \vspace{-30pt}
    
    \begin{subfigure}{0.32\columnwidth}
        \centering
        \includegraphics[width=\linewidth]{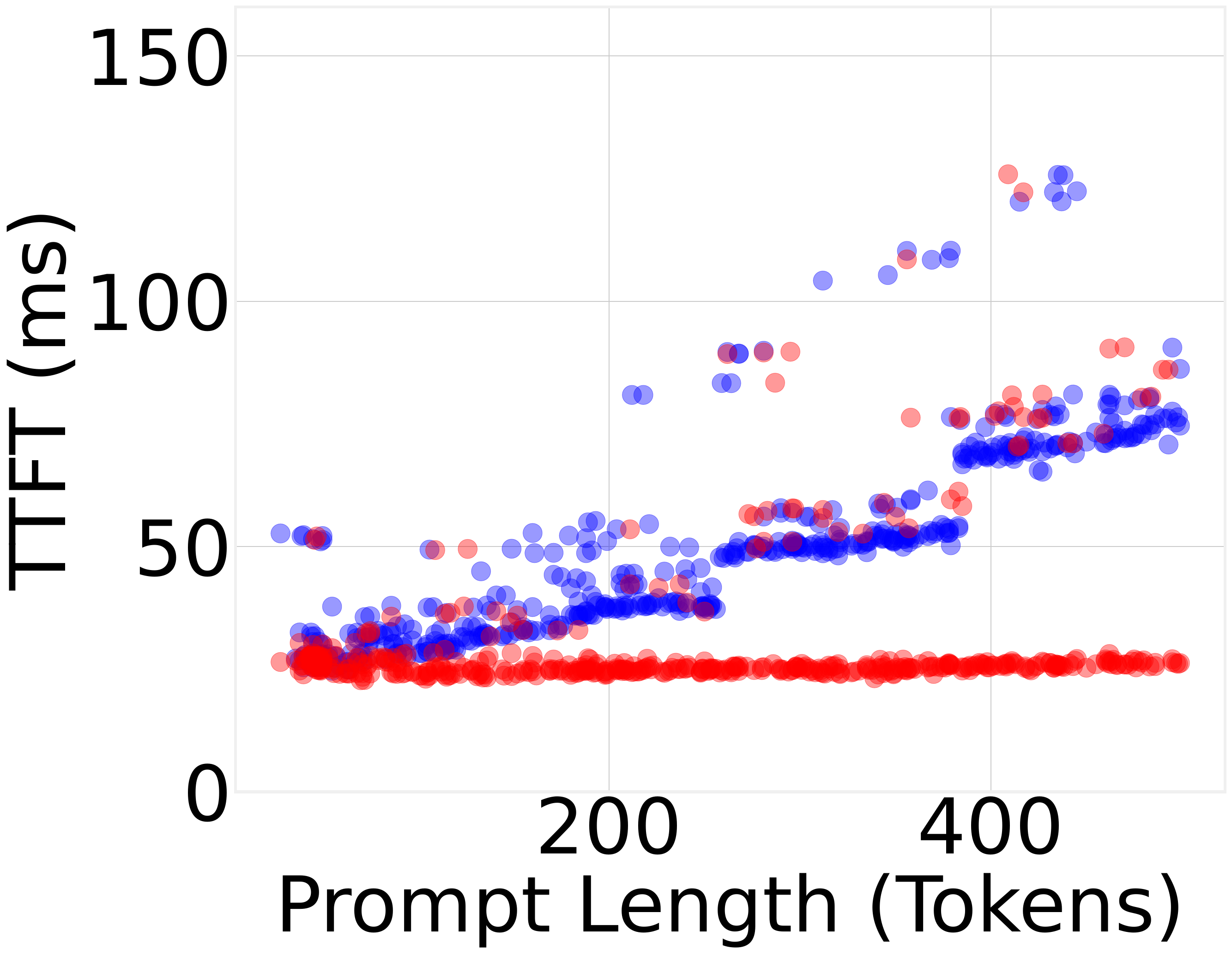}
        \caption{RPS = 10}
        \label{fig:channel_rps_10}
        \label{fig:rps10}
    \end{subfigure}
   % \begin{subfigure}{0.32\columnwidth}
    %    \centering
     %   \includegraphics[width=\linewidth]{sections/plots/ttft_linear_classifier_llama13_RPS_20.pdf}
     %   \caption{RPS = 20}
      %  \label{fig:rps20}
    %\end{subfigure}
    %\begin{subfigure}{0.33\columnwidth}
    %    \centering
    %    \includegraphics[width=\linewidth]{sections/plots/ttft_linear_classifier_llama13_RPS_30.pdf}
    %    \caption{RPS = 30}
    %    \label{fig:rps30}
    %\end{subfigure}
    \begin{subfigure}{0.32\columnwidth}
        \centering
    \includegraphics[width=\linewidth]{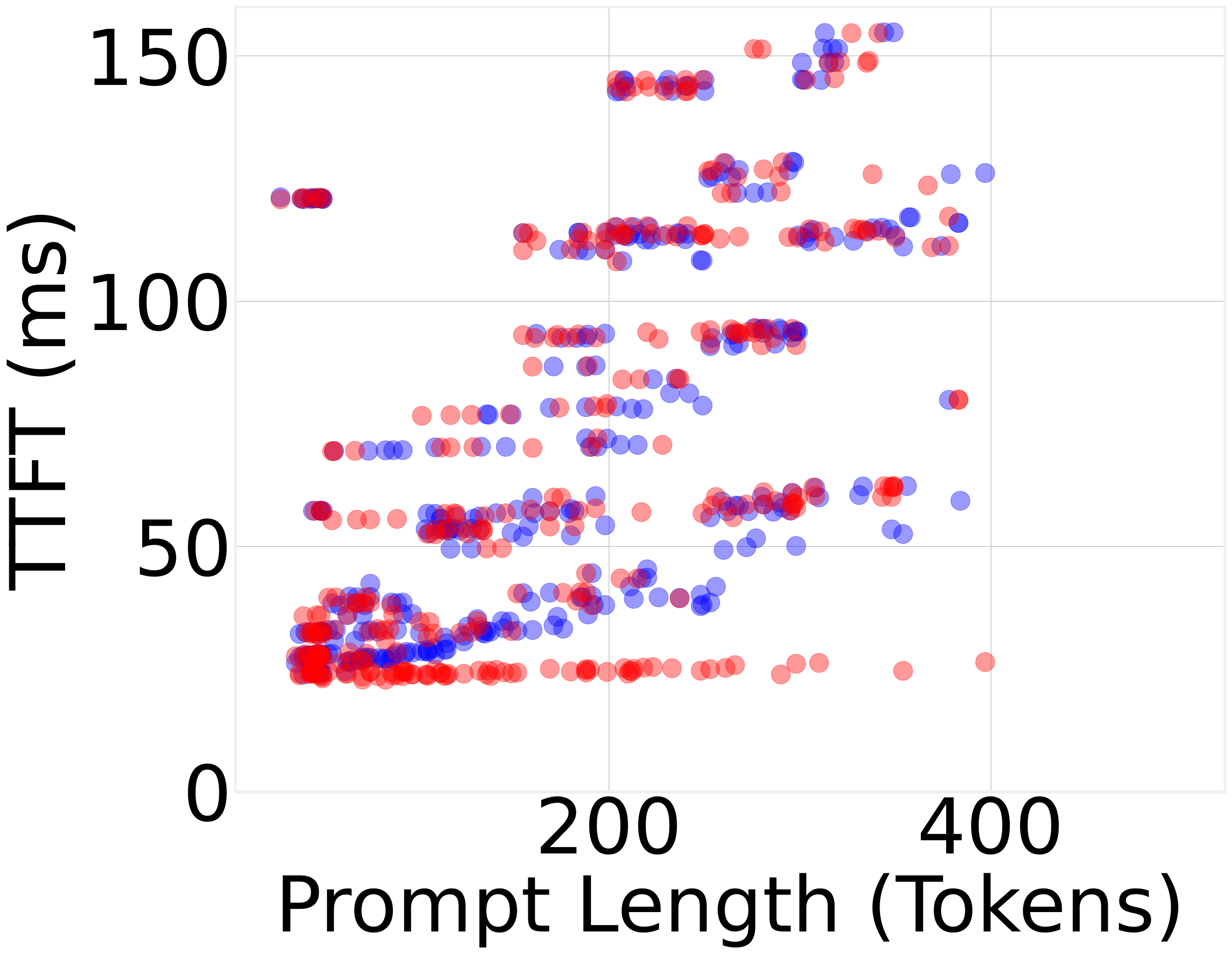}
        \caption{RPS = 40}
        \label{fig:channel_rps_40}
    \end{subfigure}
    \begin{subfigure}{0.32\columnwidth}
        \centering
        \includegraphics[width=0.9\linewidth]{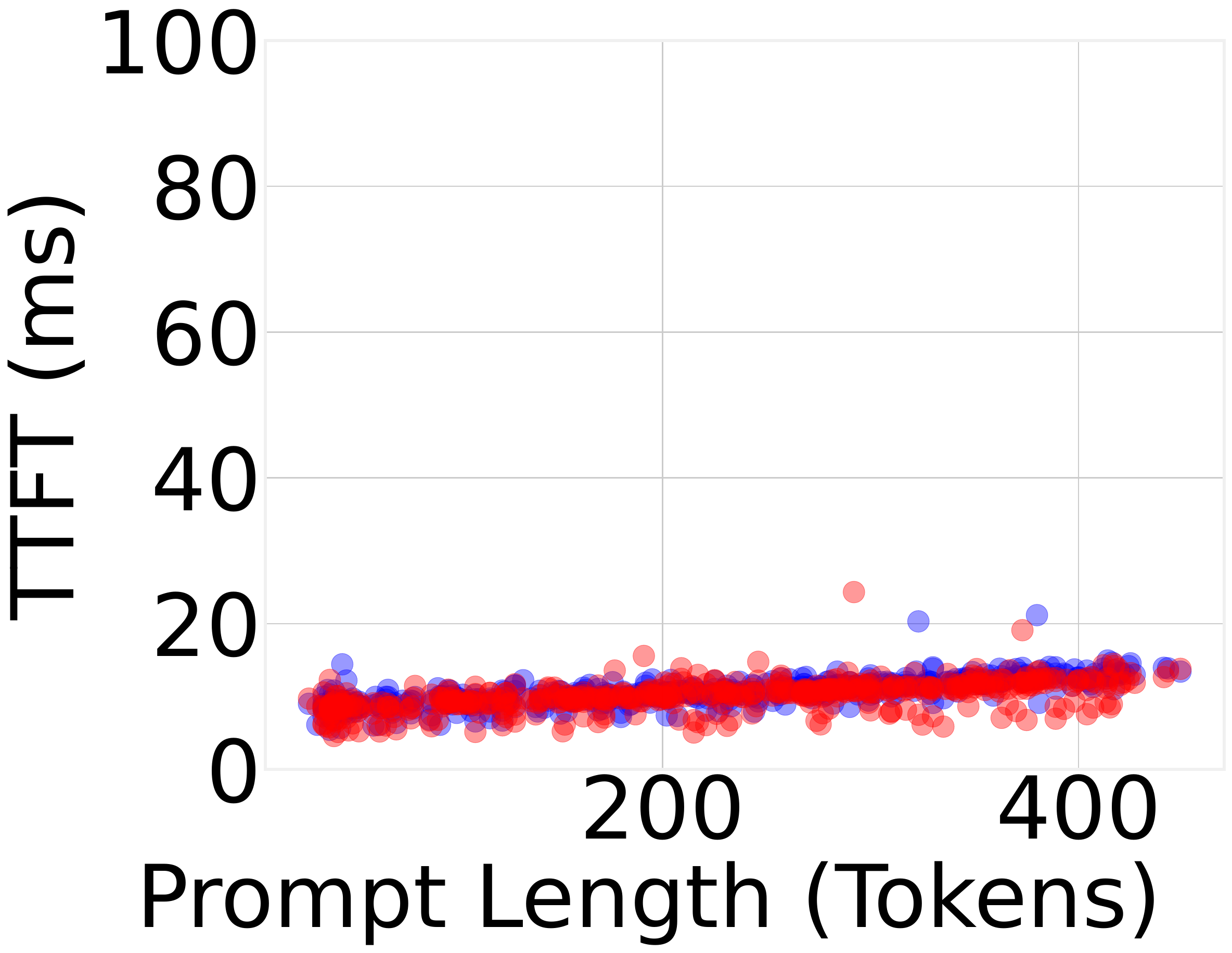}
        \caption{LLava-0.5B (A100)}
        \label{fig:channel_0.5b}
    \end{subfigure}
    \begin{subfigure}{0.32\columnwidth}
        \centering
        \includegraphics[width=0.9\linewidth]{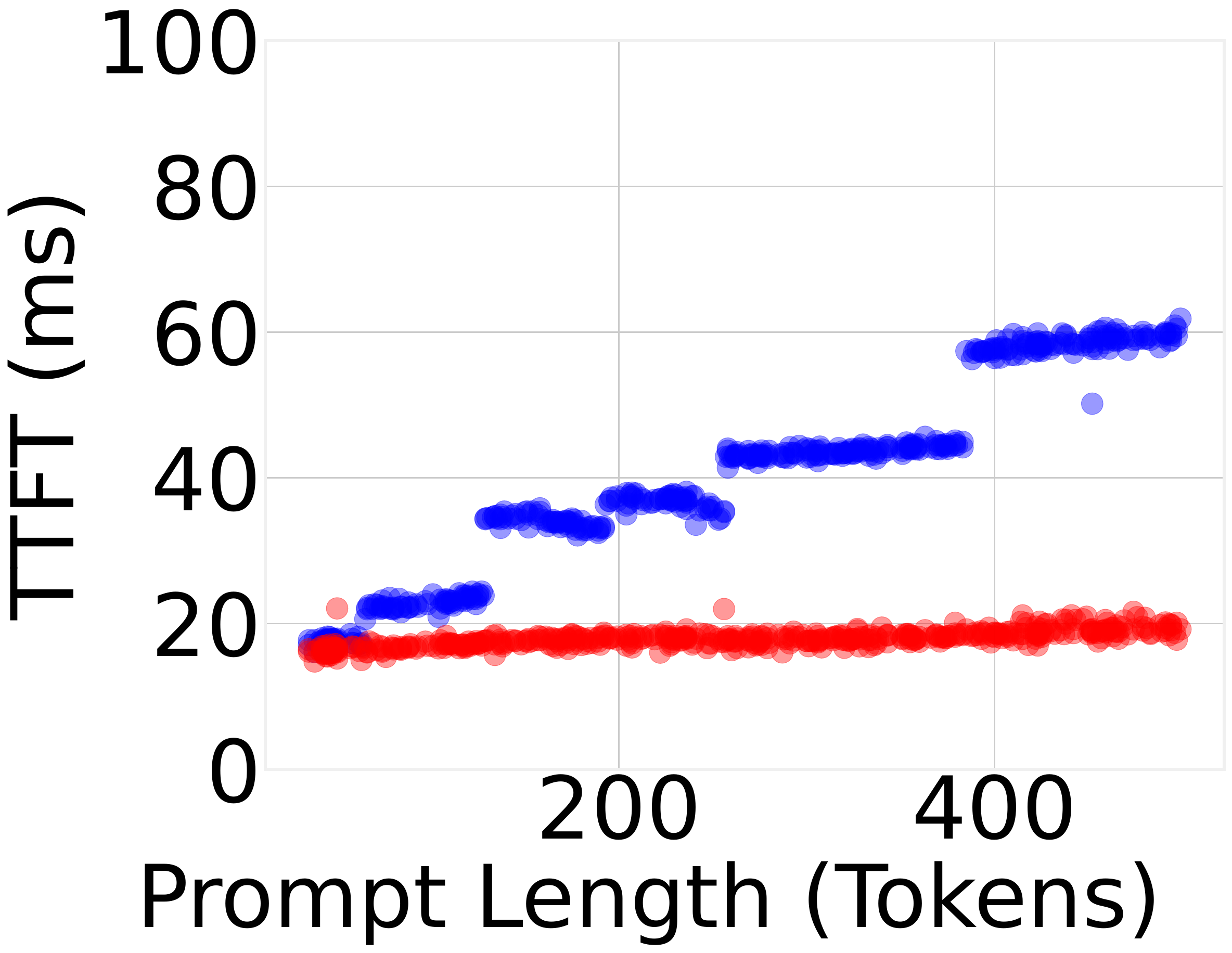}
        \caption{Llama2-7B (A100)}
        \label{fig:channel_7b}
    \end{subfigure}
    \begin{subfigure}{0.32\columnwidth}
        \centering
        \includegraphics[width=0.9\linewidth]{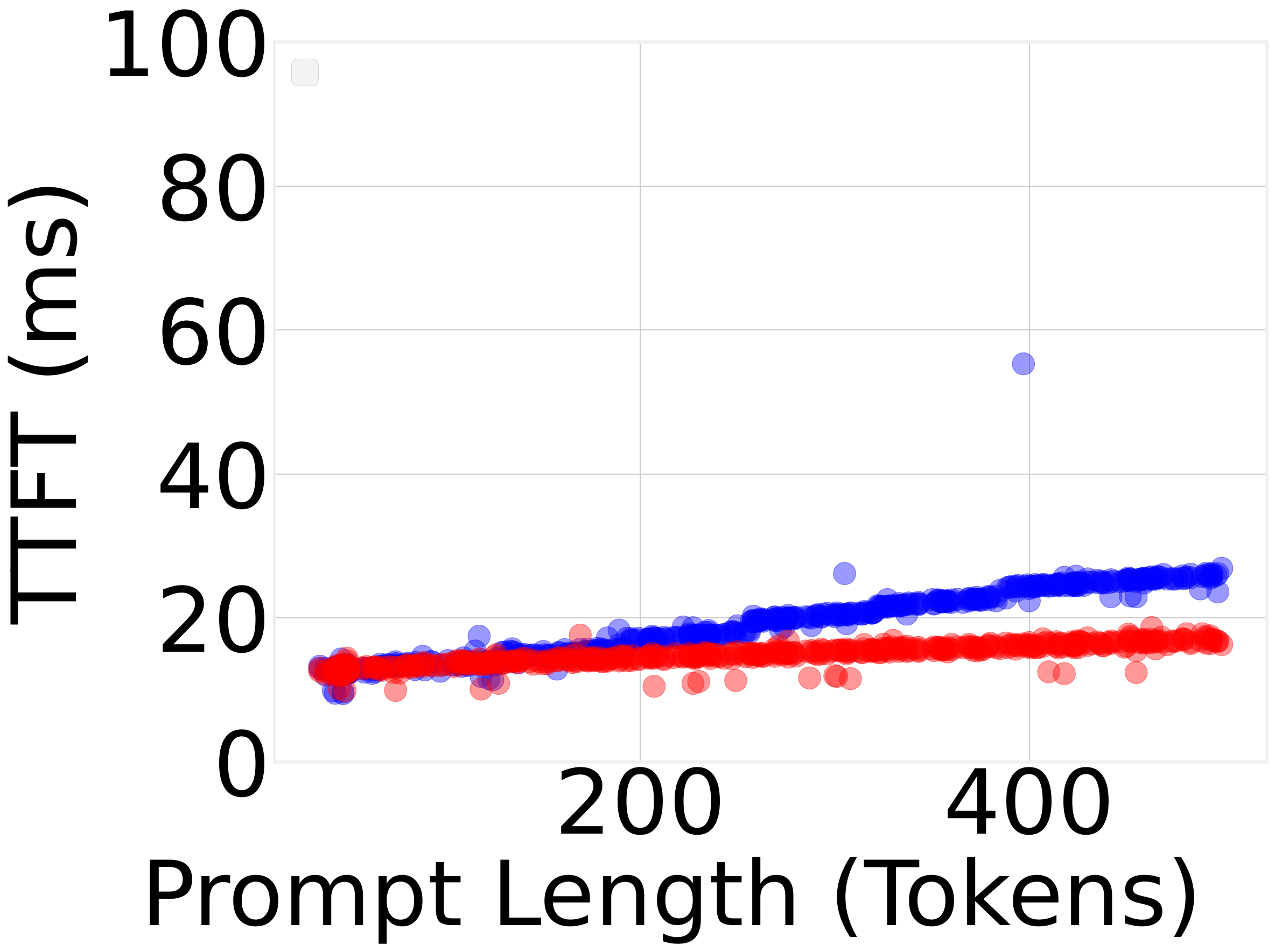}
        \caption{Llama2-7B (H100)}
        \label{fig:channel_7b_h100}
    \end{subfigure}
    \begin{subfigure}{0.32\columnwidth}
        \centering
        \includegraphics[width=0.9\linewidth]{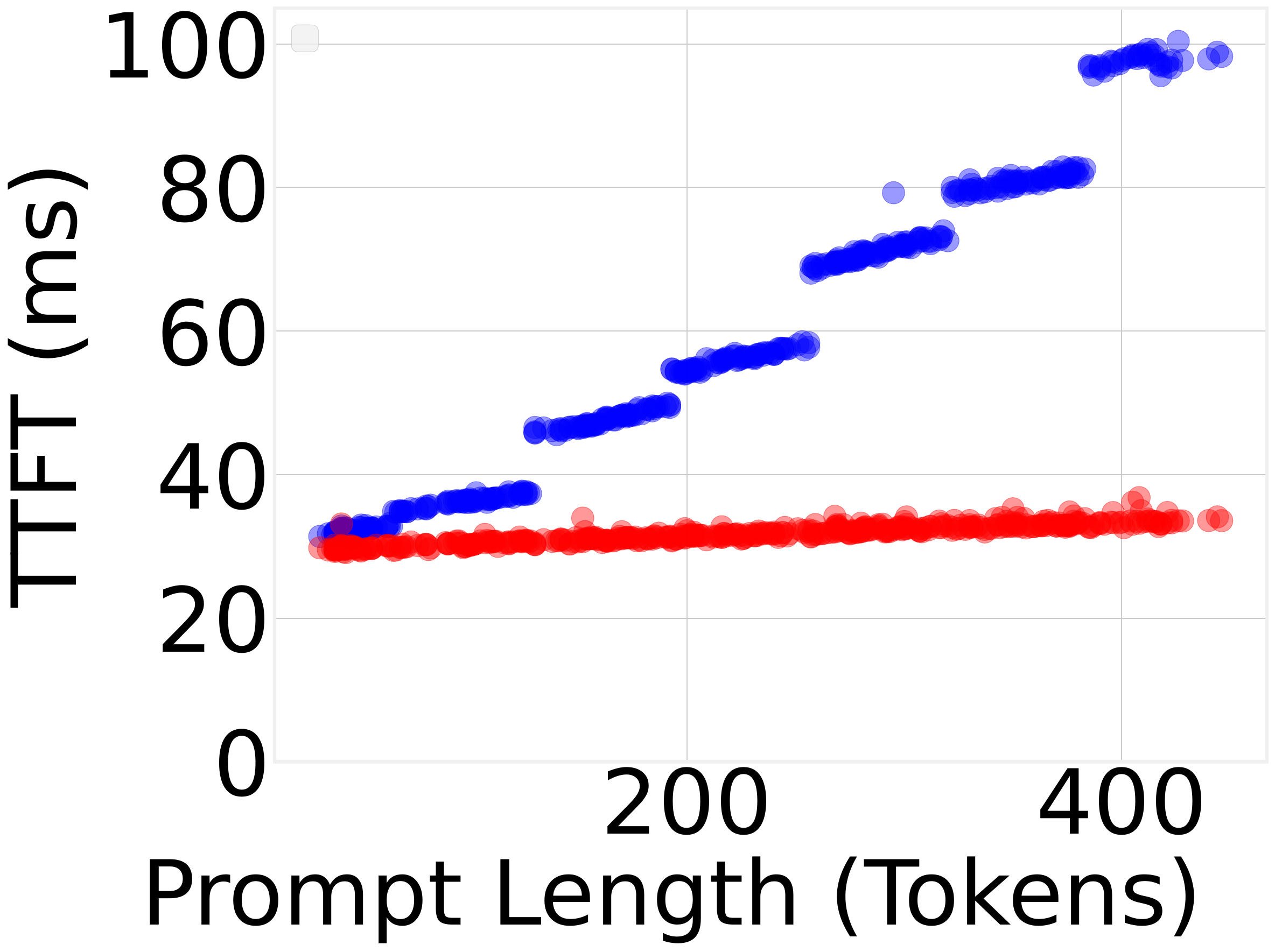}
        \caption{Llama3-70B (H100)}
        \label{fig:channel_70b}
    \end{subfigure}
\vspace{-0.1in}
    \caption{TTFT of cache hits (red) and misses (blue).}
    \label{fig:side-channels}
    \vspace{-0.1in}
\end{figure}

%%%%%%%%%%%%%%%%%%%%%%%%%%%%%%%%%%%%%%%%%%%%%%%%%%%%%%%%%%%%%%%%%%%%%%%%%%%%%%%%%%%%%%%%%%%%%%%%%%%%%%%%%%%%%%%%
\subsection{Timing Differences due to APC}\label{sec:motivation:leaks}

Although the use of the prefix cache accelerates inference, it introduces observable latency differences between cache hits and misses, particularly in the time-to-first-token (TTFT). These variations create a \emph{timing side channel}, an unintended information leak where execution time reveals whether a request shares prefixes with cached prompts. In multi-tenant environments, such leaks can be exploited by attackers to infer sensitive information. In the following examples, we illustrate how APC-induced timing differences manifest and we identify the {\it key parameters} that influence the strength and exploitability of the side channel.

\begin{example}\label{example:models-ttft-leaks} 

To illustrate APC’s impact on latency, recall that a prompt is processed as a sequence of prefixes mapped to cache entries. With APC, prefixes may either hit or miss the cache; an early miss forces recomputation of all subsequent prefixes, leading to observable differences in time-to-first-token (TTFT) between full-reuse and recomputation cases.
{Figures~\ref{fig:channel_0.5b}-\ref{fig:channel_70b}} reports TTFT comparisons across LLMs of increasing size (0.5B, 7B, 70B) and varying prompt lengths, with requests sent at a constant rate (RPS = 1). TTFT for cache hits is shown in red, while misses are shown in blue. {Llama3-70, due to its parameter count, is configured to run across 2 H100 GPUs with tensor parallelism TP = 2.} For all models except the smallest one (i.e., LLava-0.5B), we highlight the following:\looseness=-1
\begin{tightitemize}
    \item The TTFT gap between cache hits and misses become noticeable after a certain prefix length for each model.
    \item The TTFT gap grows as the shared prefix length increases.
    \item For a given prefix length, larger LLMs exhibit greater TTFT gaps.\looseness=-1
    \item Larger LLMs show distinguishable gaps even at shorter prefixes.\looseness=-1
\end{tightitemize}
For LLava-0.5B, TTFT differences between hits and misses are negligible because the model is small and the recomputation latency for the misses is minimal.

\begin{tcolorbox}[boxsep=0pt,
                  left=5pt,
                  right=5pt,
                  top=2pt,
                  bottom=2pt]
        \textbf{Observation 1:} TTFT reveals whether a request reuses cached prefixes (hits) or recomputes them (misses), and the difference becomes more pronounced with larger models and longer prefixes that are reused across users.
\end{tcolorbox}
\vspace{-0.15in}
\end{example}

\begin{figure}[t]
    \centering
    \includegraphics[width=0.9\columnwidth]{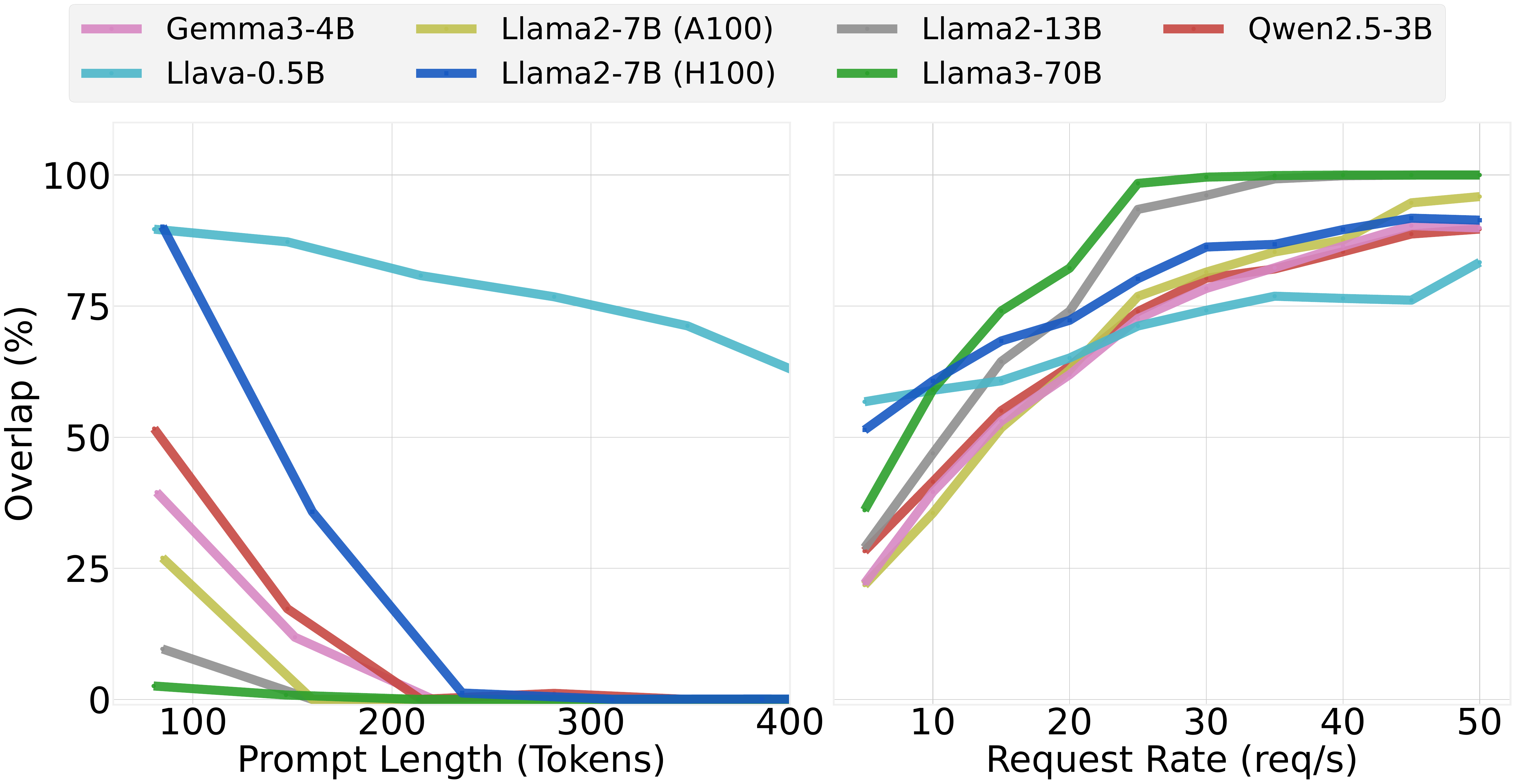}
    % \vspace{-15pt}
    \caption{Effect of the LLM model size, prefix/prompt length, system load (requests per second) and GPU hardware on the distinguishability of the APC timing differences, captured by the KDE overlap between cache hits and misses.}
    \vspace{-0.15in}
    \label{fig:KDE_models}
\end{figure}

\begin{example}\label{example:rps-ttft-leaks} 
Next, we examine how system load influences these timing differences using the same setup as before with the Llama-13B model. Figures~\ref{fig:channel_rps_10},~\ref{fig:channel_rps_40} show TTFT measurements under increasing request-per-second (RPS) rates for prompts with prefixes of up to 500 tokens. We observe that as the request-per-second (RPS) rate increases, the latency gap between cache hits and misses progressively collapses, making them hard to distinguish. This happens because higher RPS introduces resource contention and head-of-line blocking in the LLM serving pipeline~\cite{sarathi, sarathi-serve, dist-serve, konpap-euromlsys24}. Under these conditions, batching and queuing delays dominate latency, overshadowing the computational savings of prefix reuse. Thus, cache hits appear as slow as misses, effectively eliminating any timing gaps.\looseness=-1

\begin{tcolorbox}[boxsep=0pt,
                  left=5pt,
                  right=5pt,
                  top=2pt,
                  bottom=2pt]
        \textbf{Observation 2:} As system load increases, TTFT differences between prefix cache hits and misses disappear because batching and queuing delays dominate latency, masking cache-related timing variations.
\end{tcolorbox}
\end{example}
\vspace{-0.1in}
\begin{example}
{Next, we evaluate the impact of the underlying GPU hardware by running the same Llama-7B model on a single A100 vs. H100 GPU. Figures~\ref{fig:channel_7b},~\ref{fig:channel_7b_h100} show that cache misses (model inference) are served significantly faster on the H100 due to its higher throughput and improved architecture, significantly reducing timing gap compared to that of the A100.}
\begin{tcolorbox}[boxsep=0pt,
                  left=5pt,
                  right=5pt,
                  top=2pt,
                  bottom=2pt]
        {\textbf{Observation 3:} Higher-throughput GPUs accelerate inference and speed up cache misses, thereby reducing the time gap.\looseness=-1}
\end{tcolorbox}
\end{example}

To further validate the above observations across conditions, Figure~\ref{fig:KDE_models} shows the KDE overlap between the TTFT distributions of cache hits and misses across models of varying size, prefix length, system load (RPS), and {GPU hardware. All models execute on a single A100 GPU (unless otherwise noted) and Llama3-70B across 2 H100 GPUs.}
KDE overlap is a statistical measure of similarity between two probability distributions, computed as the integral of the minimum of their density functions~\cite{kde-overlap}. %In our context, it quantifies how much the TTFT distributions of cache hits and misses intersect: higher overlap indicates that hits and misses are harder to distinguish, while lower overlap means the opposite. 
Higher values indicate harder to distinguish cache hit vs. miss latency, while lower values indicate clear separation, thus stronger timing side channel.

The left plot shows that the overlap decreases for longer prefixes, especially for larger models  such as Llama2-13B and {Llama3-70B}. In contrast, %smaller models like Llava-0.5B maintain high overlap even for long prefixes, meaning that it's hard to distinguish any TTFT differences, as shown previously in Figure~\ref{fig:side-channels}. 
smaller models (e.g., Llava-0.5B) have distinctly higher overlap, { as well as models that run on newer GPUs (Llama2-7B on H100) for short prompts.}  The right plot shows that overlap increases with request rate (RPS), as batching and queuing delays dominate latency and mask cache-dependent variations. At high RPS, hits and misses become nearly indistinguishable across all models, {regardless of the underlying GPU A100 vs. H100}, effectively eliminating the timing leak. 

In conclusion, while prior work acknowledges APC-induced timing side channels~\cite{chu2026selectivekvcachesharingmitigate, cache-partitioning, input-snatch, pmlr-v267-gu25b}, our motivational analysis shows that the strength and exploitability of the channel depend on the following critical parameters: 
\begin{tightitemize}
    \item {\bf the shared prefix's length:} longer prefixes are more vulnerable.
    \item {\bf the size of the LLM model:} larger models are more vulnerable.
    \item {\bf the system load}: lower load (requests received per second) is more vulnerable.
    \item {\bf the underlying hardware platform:} since it directly influences timing measurements. 
\end{tightitemize}

These parameters jointly determine whether timing side channels are distinguishable and, should be taken into consideration into building robust and practical mitigation mechanisms.

\begin{tcolorbox}[boxsep=0pt,
                  left=5pt,
                  right=5pt,
                  top=2pt,
                  bottom=2pt]
        \textbf{Key insight:} While APC introduces timing side channels, their exploitability depends on various parameters: the prefix length, LLM size, system load, and hardware platform. Prior work treated APC leaks as uniformly exploitable, whereas our analysis reveals that these parameters jointly determine side-channel strength and must guide efficient and practical mitigation.
\end{tcolorbox}

% In multi-tenant scenarios, these leaks can be exploited to do prompt-stealing attacks
%% Give high-level idea of how prompt-stealing attacks using APC-leaks work
%% Give data about prefix-sharing in established workloads

%As shown above, requests that shares a common prefix with prior requests are faster than requests that do not share prefixes. 
%%%%%%%%%%%%%%%%%%%%%%%%%%%%%%%%%%%%%%%%%%%%%%%%%%%%%%%%%%%%%%%%%%%%%%%%%%%%%%%%%%%%%%%%%%%%%%%%%%%%%%%%%%%%%%%%
\subsection{Prompt Stealing Attacks via APC leaks}
\label{sec:motivation:attacks}

%%%%%%%%%%%%%%%% attack description %%%%%%%%%%%%%%%%%%%
%
In multi-tenant systems the prefix cache is shared between requests from different users and security domains.
Hence, these visible timing differences can be exploited by attackers to incrementally reconstruct 
sensitive prompts or steal private information (secrets) of other users by observing the TTFT of attacker requests.
That is, APC leaks can be exploited to perform \emph{prompt stealing attacks}~\cite{early-bird}.
Figure~\ref{fig:attack} illustrates the workflow of a timing-based prompt stealing attack that exploits APC leaks, inspired by~\cite{early-bird}.
First, the victim user issues a request that includes a prompt containing sensitive information.
%
%For simplicity, let's assume that the prompt follows a fixed template (known by the attacker) and it includes user-specific secrets.
%
%For example, a prompt template can be {\it Compose a meeting agenda for an interdisciplinary team discussing the treatment plan for [Name] with [medical condition]}), where {\it [Name]} and {\it [medical condition]} are secret.
%
The attacker uses an LLM-based \emph{Prompt Constructor} to create candidate prompts. %by varying words within the template.
Next, each candidate is submitted to the LLM serving system, and the TTFT is measured.
By observing the TTFT, the attacker can determine (e.g., using a dedicated pre-trained classifier) whether the observed latency corresponds to hits or misses in the prefix cache. 
In particular, a cache hit indicates that the candidate shares a prefix with the victim’s prompt. 
By repeating this process iteratively, the attacker can reconstruct the original prompt word-by-word or retrieve missing/private data fields (e.g., names, dates, identifiers) without direct access to the cache~\cite{early-bird, pmlr-v267-gu25b}.
%\end{example}

\smallskip
\noindent
{{\bf Attack Examples.}
Consider an LLM-based customer-support application where multiple users send requests to a shared backend model. To improve performance, the system prepends hidden context (e.g., account data) to each query and reuses cached prefixes across requests. For example, the system may internally construct a prompt such as: \textit{“You are a support assistant. Customer name: John Doe. Account ID: 48291.”}, followed by \textit{“Summarize recent activity.”} An attacker, acting as a regular API user, can issue inputs containing guesses for likely fields (e.g., “Customer name: A…”, “Customer name: Alice…”) and measure response times. Faster responses indicate prefix reuse, allowing the attacker to iteratively refine guesses and recover sensitive fields embedded in prior prompts.}

{
Next, consider a shared LLM API where multiple users submit prompts that may include sensitive personal or business information. For example, a user might send a query such as \textit{“Draft an email to my manager John Doe about delaying Project Phoenix.”} As requests are processed, parts of these prompts may be cached to improve performance. An attacker, interacting as a normal user, can probe the system by issuing prompts containing guesses (e.g., names or project titles) and measuring response times. By iteratively refining these guesses (e.g., extending “John…” to “John Doe…”), the attacker can reconstruct real user prompts.
}

%%%%%%%%%%%%%%%% real secrets %%%%%%%%%%%%%%%%%%%
\smallskip\noindent
{{\bf Real prompts with secret information.} Next, in Table~\ref{tab:secret-datasets} we analyze datasets with real user prompts from diverse domains, including logs of personal chats, interactions with assistants, and instruction tuning; the task where a prompt specifies an instruction and the model is trained to produce the intended response. Given that these datasets do not include labeled private information, we use the {\tt ai4privacy 0.5.0} Python module~\cite{ai4privacy_pypi_v050} as a proxy to identify sensitive content within each prompt and estimate the fraction of prompts containing personally identifiable information (PII), as also done in prior work on LLM privacy leakage~\cite{ai4privacy2024,cheng2025pii,singh2025unmasking,acharya2025comparative}.}
%Its transformer-based named entity recognition architecture, broad entity coverage across dozens of PII categories, and open availability on HuggingFace make it a reproducible and well-validated choice for auditing PII prevalence in large text datasets.} 

{We observe that a substantial fraction of prompts across datasets include sensitive content, ranging from 19\%-27.24\% in user conversations and assistant interactions, up to 76.1\% in instruction-tuning prompts. This gap arises because instruction-tuning prompts ground tasks in concrete contexts, often including specific entities such as names, locations, or dates. For example, a prompt may ask: \textit{``Summarize the following article about Barack Obama’s visit to Berlin in 2013''} or \textit{``Extract all locations and dates mentioned in a passage about the 2020 Tokyo Olympics,''} which naturally include personally identifiable information. In contrast, conversational prompts are typically more abstract or task-oriented, e.g., \textit{``How do I improve my resume?''} or \textit{``What’s a good itinerary for a 3-day trip?''}, and thus rarely include personal details.
}

{
We further evaluate the extent to which such information can be recovered using a state-of-the-art BERT-based model~\cite{devlin2019bert} (google-bert/bert-base-uncased), mimicking an attacker employing a Prompt Constructor. We find that, among prompts containing secrets, the model can recover hidden information in up to 52.5\% of cases. This result indicates that real-world prompt datasets contain substantial amounts of recoverable sensitive information, suggesting that prompt stealing attacks can be effective in extracting real secrets.
}

\begin{table}[t]
\centering
\small
\begin{tabular}{llccc}
\toprule
\textbf{Dataset} & \textbf{Domain} & \textbf{\shortstack{Prompts}} & \textbf{\shortstack{\% Prompts \\ w/ Secrets}} & \textbf{\shortstack{\% Secrets \\ Recover.}} \\
\midrule
Flan~\cite{dataprovenance2024commercialflan_dialog} & Instr. tuning & 553,869 & 76.1\%  & 40.39\% \\
UltraChat~\cite{ultrachat_dataset} & Chat logs & 10,000 & 27.24\% & 52.5\% \\
ShareGPT~\cite{anon8231489123_2023} & Chat logs & 702,151 & 24.3\% & 22.8\% \\
OASST1~\cite{openassistant_oasst1} & Assistant logs & 39,283 & 19\% & 30.76\% \\
\bottomrule
\end{tabular}
\caption{{Sensitive information (secrets) in real-world prompt datasets and its recovery rate.} }
\label{tab:secret-datasets}
\vspace{-0.3in}
\end{table}

{In addition, }prior studies~\cite{chu2026selectivekvcachesharingmitigate} have shown that such real world datasets %such as ShareGPT~\cite{anon8231489123_2023}, Multiturn Chat~\cite{bellegroup_2023}, and Prompt Multitasks~\cite{gallego_2024}, 
exhibit frequent prefix reuse within and across users. Specifically, requests from the same user share prefixes in 9–60\% of cases, while up to 30\% reuse occurs across different users. {This widespread reuse increases the likelihood that sensitive prompt components remain in cache and become vulnerable to prompt stealing attacks.}

%%%%%%%%%%%%%%%%%%%%%%%%%%%%%%%%%%%%%%%%%%%%%%%%%%%%%%%%%%%%%%%%%%%%%%%%%%%%%%%%%%%%%%%%%%%%%%%%%%%%%%%%%%%%%%%%
\subsection{Defenses Against APC leaks}\label{sec:motivation:defenses}

Existing defenses against APC-induced timing leaks fall into three main categories, each with distinct trade-offs between security and performance, as summarized in Table~\ref{tab:related-works}.

% Recent works have proposed several defenses against APC-induced timing leaks, which fall into three main categories, each with distinct trade-offs between security and performance, as summarized in Table~\ref{tab:related-works}.

\paragraph{Cache Isolation and Partitioning.}  
Several works propose mitigating APC leaks by  eliminating cache sharing entirely across different users and security domains.
InputSnatch~\cite{input-snatch} and Auditing Prompt Caching~\cite{pmlr-v267-gu25b} advocate per-user cache isolation or disabling prefix reuse across requests.
Cache Partitioning~\cite{cache-partitioning} similarly enforces user-boundaries within the KV-cache to prevent cross-user hits. 
{At the same time, major AI providers adopt isolation mechanisms in-production at varying scopes, along with periodic cache refresh policies, as summarized in Table~\ref{tab:provider-mitigations}. OpenAI scopes its prompt cache at the organization level and explicitly warns about timing observability~\cite{openai2024promptcaching}. Anthropic recently tightened its isolation boundary from the organization to the workspace level~\cite{anthropic_caching_docs}. Similarly, Azure OpenAI and Amazon Bedrock enforce subscription- and account-level isolation, largely driven by compliance requirements~\cite{azure_openai_docs, amazon_bedrock_docs}. Google Vertex AI provides manual caching primitives with per-project scoping~\cite{google_vertex_caching}. 
In the open-source ecosystem, the widely used vLLM framework introduced an optional cache isolation mechanism based on a per-request \texttt{cache\_salt}~\cite{vllm_cache_salt_pr}, following the disclosure of CVE-2025-46570~\cite{cve202546570} (a publicly registered security vulnerability) and the related GitHub Security Advisory (GHSA-4qjh-9fv9-r85r), which highlighted the risk of timing leakage from shared prefix caches. Finally, cache refresh times range from few minutes up to hours and days, leaving ample time for an attacker to mount prompt stealing attacks targeting cached prefixes.}

\begin{table}[t]
\centering
\small
\begin{tabular}{|l|c|p{1cm}|p{1cm}|}
\hline
\textbf{Related Work} & \textbf{Mitigation} & \textbf{Prefix Reuse} & \textbf{Low TTFT} \\\hline
Partitioning~\cite{cache-partitioning} & Per-user Isolation & \xmark & \xmark \\\hline
InputSnatch~\cite{input-snatch} & Per-user Isolation & \xmark & \xmark \\\hline
Audit~\cite{pmlr-v267-gu25b} & Per-user Isolation & \xmark & \xmark \\\hline 
Remote~\cite{remote-attacks} & Timing Obfuscation& \cmark & \xmark \\\hline
% Early Bird~\cite{early-bird} & Selective Sharing & \cmark & \xmark \\\hline
SafeKV~\cite{chu2026selectivekvcachesharingmitigate} & Selective Sharing & \cmark & \xmark \\\hline
\hline
\textbf{\sys} & Selective Sharing & \cmark & \cmark \\\hline
\end{tabular}
\caption{Comparison of current mitigation strategies.}
\label{tab:related-works}
\vspace{-0.2in}
\end{table}

{
%Cache-based timing
APC-based
side channels are a practical and widely acknowledged threat across major AI providers.} 
Cache isolation  ensures strong security guarantees by preventing cross-user prefix sharing.
% prevent timing leaks due to shared prefixes by addressing the leak at its source.
%
However, it also sacrifice the performance benefits of prefix caching, which can result in significant overhead given that ``real-world LLM workloads frequently exhibit significant cross-query reuse''~\cite{chu2026selectivekvcachesharingmitigate}.\looseness=-1

\begin{tcolorbox}[boxsep=0pt,
                  left=5pt,
                  right=5pt,
                  top=2pt,
                  bottom=2pt]
        \textbf{Shortcoming 1:} User-based cache isolation prevents timing leaks but introduces overhead due to reduced prefix sharing.
\end{tcolorbox}

\paragraph{Timing Obfuscation.}  
%
%On the other extreme, 
\citeaut{remote-attacks} propose to secure systems by making the execution time of inference ``constant'' across hits and misses.
For this, they propose to inject noise in the system to mask timing signals and, ultimately, close the side channel by making hits and misses in the prefix cache indistinguishable to attackers.
That is, these defenses would ensure that the distributions of hits and misses in figure~\ref{fig:side-channels} overlap.
Although obfuscation defenses can prevent APC leaks, they have two limitations.
First, they introduce uniform delays across all requests, penalizing benign users and reducing overall performance.
Second, they do not address the source of the timing leak (i.e., the prefix cache being shared between security domains) and historically obfuscation defenses for timing leaks are less robust since attackers can often find ways of amplifying even small differences in execution time~\cite{browser1,10.5555/3241094.3241131,gras_aslr_2017}.

\begin{tcolorbox}[boxsep=0pt,
                  left=5pt,
                  right=5pt,
                  top=2pt,
                  bottom=2pt]
        \textbf{Shortcoming 2:} Obfuscation-based techniques do not address the source of leaks and introduce overheads  for benign requests.
\end{tcolorbox}

\begin{table}[t]
\centering
\small
\begin{tabular}{lll}
\toprule
\textbf{Provider} & \textbf{Isolation Scope} & \textbf{Cache Refresh} \\
\midrule
OpenAI~\cite{openai2024promptcaching}          & Per organization   & 5--60 min \\
Anthropic~\cite{anthropic_caching_docs}    & Per workspace%\textsuperscript{$\dagger$} 
& Session only (volatile) \\
Azure OpenAI~\cite{azure_openai_docs}      & Per subscription   & Not disclosed \\
Amazon Bedrock~\cite{amazon_bedrock_docs}  & Per account/org    & 5 min \\
Google Vertex~\cite{google_vertex_caching} & Per GCP project    & 1 min--30 days (TTL) \\
vLLM~\cite{vllm_cache_salt_pr,vllm_cve_2025} & None by default%\textsuperscript{$\ddagger$} 
& Unbounded \\
\bottomrule
%\multicolumn{3}{l}{\textsuperscript{$\dagger$}Tightened from org-level in Feb.~2026.} \\
%\multicolumn{3}{l}{\textsuperscript{$\ddagger$}Opt-in isolation added after CVE-2025-25183 and GHSA-4qjh-9fv9-r85r.} \\
\end{tabular}
\caption{Cache isolation across LLM providers.}
\label{tab:provider-mitigations}
\vspace{-0.3in}
\end{table}

\paragraph{Selective Cache Sharing.}  
To balance security and efficiency, some defenses restrict cache sharing rather than disabling it completely.
%
% \citeaut{early-bird} propose limiting sharing to prefixes of at least $k$ tokens, reducing leakage granularity (without fully solving the problem) but still degrading performance. 
%
% In contrast, 
SafeKV~\cite{chu2026selectivekvcachesharingmitigate} introduces a multi-tier detection pipeline that semantically classifies requests using rule-based checks and semantic validation via LLMs either as ``sensitive'' (i.e., that may contain secret information) or non-sensitive.
In SafeKV, only requests classified as sensitive are isolated at cache-level, thereby preventing other users (attackers included) from hitting on them.
Although this approach prevents leakage without fully disabling APC, its reliance on heavyweight semantic LLM-based analysis limits scalability for high-performance deployments.
Moreover, some requests might be misclassified and, therefore, left unprotected.

\begin{tcolorbox}[boxsep=0pt,
                  left=5pt,
                  right=5pt,
                  top=2pt,
                  bottom=2pt]
        \textbf{Shortcoming 3:} Current selective sharing defenses may fail in preventing leaks for ``misclassified'' requests and still introduce unnecessary overhead.
\end{tcolorbox}

These shortcomings motivate the need for a defense that adapts to the conditions under which timing leaks are exploitable, secures prefix reuse without blind per-user isolation, and preserves the efficiency of APC across diverse workloads, without introducing unbearable overheads in return for security.

\section{System Design}
\label{sec:system}

In this section, %we introduce \sys{}, that provides system-level security against timing leaks induced by APC.
we start by providing a high-level overview of \sys{} and its objectives (\ref{sec:system-overview}),
and later describe \sys{}'s core components (Sections~\ref{sec:system-detector} and~\ref{sec:system-load-monitor}).

%%%%%%%%%%%%%%%%%%%%%%%%%%%%%%%%%%%%%%%%%%%%%%%%%%%%%%%%%%%%%%%%%%%%%%%%%%%%%%%%%%%%%%%%%%%%%%%%%%%
\subsection{Overview and Objectives}
\label{sec:system-overview}

\textbf{\sys} deploys a lightweight defense mechanism that secures multi-tenant LLM serving systems against timing side-channel attacks caused by Automatic Prefix Caching (APC). 
It builds on the \textbf{key insight} that preventing timing side-channel attacks does not require isolating entire users (like it is done in~\cite{cache-partitioning,input-snatch,pmlr-v267-gu25b}); it only requires isolating \emph{prefixes} that might lead to attacks. 
Building on this insight, \sys{} allows cached prefixes to be reused across users but stops reuse beyond prefixes that could reveal sensitive information.
This selective prefix isolation works on a simple principle: attacks require different users to share and reuse a prefix. \sys{} enforces this by tracking how users interact with shared cache entries 
%Our selective-isolation criterion is built on top of a necessary pre-requisite of attacks (i.e., different user requests must share a prefix) and it relies solely on monitoring user interactions with shared cache entries, 
without requiring any knowledge of prompt semantics~\cite{chu2026selectivekvcachesharingmitigate}.
% The decision to enforce selective sharing relies solely on monitoring user interactions with shared cache entries, without requiring any knowledge of prompt semantics~\cite{safeKV}. 
This approach embodies the principle of cooperative efficiency: benign users benefit from shared caching through \textbf{solidarity}, where common prefixes remain accessible to all. 
When suspicious behavior threatens this solidarity, \sys intervenes by isolating the corresponding prefixes rather than penalizing all users. 
This design ensures that cache reuse within and across users is maximized to the greatest extent possible without compromising security.

\vspace{0.05in}
\noindent{\it Objectives.}
The system design is guided by the following goals:
\begin{itemize}[leftmargin=25pt, itemsep=0pt, topsep=0pt]
    \item[\textbf{[O1]}] \textbf{Secure Prefix Reuse:} Prevent timing attacks introduced by Automatic Prefix Caching (APC) by isolating suspicious prefixes, avoiding heavy-handed approaches such as per-user isolation or disabling cache sharing.
    
    \item[\textbf{[O2]}] \textbf{Performance Preservation:} Maximize cache reuse and accelerate inference speed, across diverse workloads and LLM models, thereby ensuring that security does not come at the cost of performance.
    
    \item[\textbf{[O3]}] \textbf{Lightweight and Practical Design:} Provide real-time protection with minimal overhead, without requiring semantic analysis or prior knowledge of sensitive content, and integrate seamlessly with state-of-the-art LLM serving systems.
\end{itemize}

\begin{figure}[t]
    \centering
    \includegraphics[width=1.1\columnwidth]{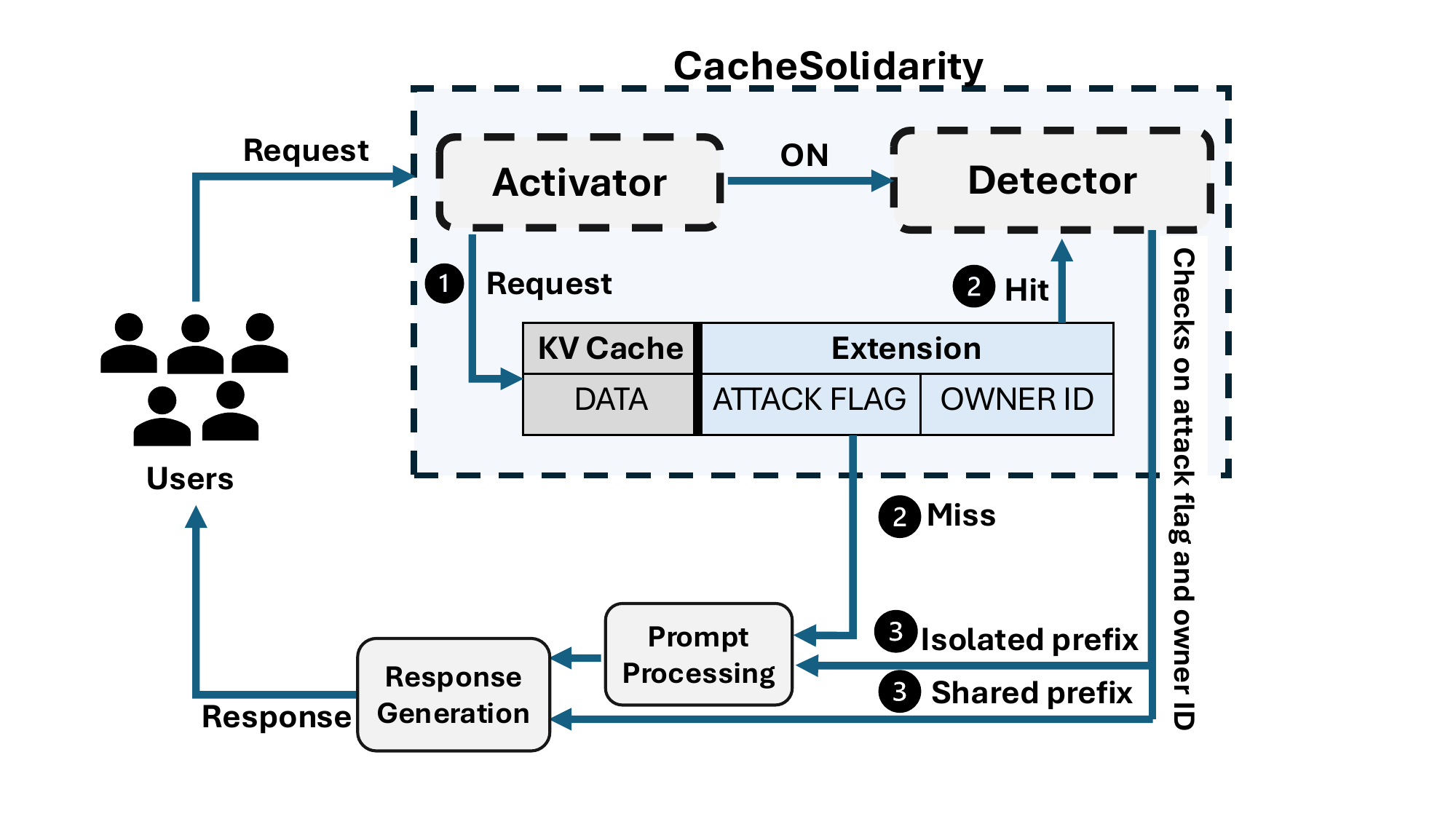}
    \vspace{-0.3in}
    \caption{{System Design of \sys.}}
    \vspace{-0.15in}
    \label{fig:system}
\end{figure}

\vspace{0.05in}
\noindent{\it Overview.}
To achieve these objectives, \sys introduces a lightweight pipeline that operates on the request-serving path, as illustrated in Figure~\ref{fig:system}. It extends the KV cache with minimal metadata per entry %to enable fine-grained tracking of user-prompt interactions and flag suspicious prefixes that are shared and reused across users. 
to track cross-user prefix reuse and flag suspicious entries. When a request arrives, the \emph{Activator} %first decides whether selective isolation should be enabled based on the distinguishability of the timing side channel, under the current LLM model, prefix length, system load and underlying hardware platform. 
determines whether to enable the \emph{Detector} based on the distinguishability of timing differences in past cache hits and misses. At the same time, the LLM serving system checks %whether the prefixes that are part of the request hit or miss the cache. 
if the request's prefixes hit or miss the cache. On a cache miss, a new cache entry is created and tagged with the user’s ID. %; no further action is needed because misses do not leak timing information.
On a cache hit, the \emph{Detector} (if enabled by the \emph{Activator}) checks whether the prefix can be safely reused.
%determines whether the prefix should be isolated to prevent attacks or shared as usual, following the procedure described in Section~\ref{sec:system-detector}. 
%If the prefix is shared (marked as 3 in Figure~\ref{fig:system}), the request proceeds with full reuse and the response is returned directly to the user. If the prefix is isolated, \sys{} stops reuse at that point and recomputes the remaining tokens through the LLM, creating a new cache entry for the isolated prefix. This results in a \emph{partial hit}: the request benefits from reuse up to the shared prefix, while the remaining portion is isolated and recomputed to prevent leakage.
If safe, the request proceeds with full reuse and is forwarded to the LLM for decoding. Otherwise, a {\it partial hit} occurs: reuse stops at the flagged prefix, and the remaining prefill tokens of the isolated prefix are recomputed before proceeding to the LLM for decoding.

\vspace{0.05in}
\noindent{\it System Components.} \sys consists of the following:

\begin{tightitemize}
    \item \textbf{KV Cache Extension:} Augments each cache entry with metadata to track user-prompt interactions and flag suspicious prefixes (\S\ref{sec:system-detector}).\looseness=-1
    \item \textbf{Detector:} Examines prefix reuse to identify cross-user access and enforce selective isolation when suspicious reuse occurs (\S\ref{sec:system-detector}).
    \item \textbf{Activator:} Dynamically activates selective isolation only under conditions that create exploitable timing side-channels, balancing security with performance (\S\ref{sec:system-load-monitor}).
\end{tightitemize}
\noindent
Together, these components maximize cache reuse while mitigating timing attacks without sacrificing performance.
%Section~\ref{sec:system-example} illustrates an example worklflow of \sys's selective prefix sharing. Section~\ref{sec:system-sec-analysis} discusses the security analysis and potential limitations of \sys.

%%%%%%%%%%%%%%%%%%%%%%%%%%%%%%%%%%%%%%%%%%%%%%%%%%%%%%%%%%%%%%%%%%%%%%%%%%%%%%%%%%%%%%%%%%%%%%%%%%%
\subsection{KV Cache Extension and Detector}
\label{sec:system-detector}

\subsubsection{KV Cache Extension}
\sys extends each cache entry with the following metadata fields:

\begin{tightitemize}
  \item \textbf{OwnerID}: the identifier of the user who first created and populated this cache entry. It is set exactly once at allocation time and remains immutable.
  \item \textbf{AttackFlag}: a marker indicating that this prefix has been flagged for isolation due to being reused by multiple users. When set, prefix reuse beyond this point must be disabled for non-owners.
\end{tightitemize}
\noindent
These fields add only a few bytes per entry and do not affect tensor layout or memory mapping.
%

%Next, we explain why we track these metadata.

%\paragraph{Why Track the Owner?}
%Tracking \texttt{OwnerID} is essential because any user other than the original creator is a potential attacker and may attempt to infer information about private fields of a prefix by exploiting timing differences. By recording ownership, \sys can distinguish benign reuse from suspicious one and enforce selective isolation only when necessary.

%\paragraph{Why Flag a Cache Entry?}
%Flagging an entry is essential because cross-user reuse is the point where timing side channels become exploitable. When a request reuses cache entries created by another user, the risk of prompt stealing begins. By flagging a suspicious cache entry, \sys sets a boundary: reuse up to this point is allowed, but going further is restricted for non-owners (potential attackers).

\subsubsection{Detection Pipeline}
When a cache entry is created on a cache miss, its metadata is initialized: \texttt{OwnerID} is set to the current user ID and \texttt{AttackFlag} is cleared. When there is a cache hit, the request is sent to the \emph{Detector} for validation. At this stage, the Detector examines whether the prefix should remain shared or be isolated to prevent leakage. Two scenarios can occur during validation:
\begin{tightitemize}
    \item \textbf{Hit on an unflagged prefix:} If the prefix is unflagged and owned by the requesting user, reuse proceeds without changes, the cached entry is returned back to the user. If the prefix is owned by a different user, the Detector flags this prefix as isolated for future requests, ensuring that reuse beyond this point is blocked for non-owners.
    \item \textbf{Hit on a flagged prefix:} If the prefix is previously flagged, the Detector checks whether the next prefix in the prompt belongs to the requesting user. If yes, reuse continues normally. If not, reuse stops at the flagged prefix, and the remaining ones are recomputed through the LLM.
\end{tightitemize}

\begin{figure*}[t]
    \centering
    \includegraphics[width=\textwidth,keepaspectratio]{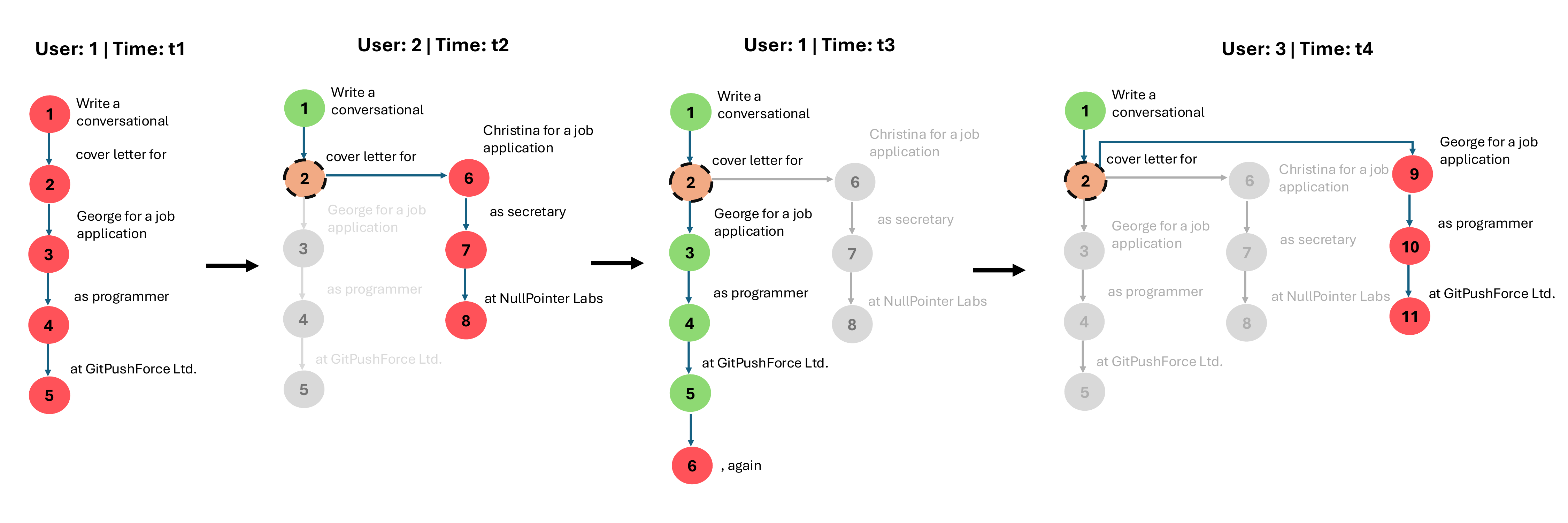}
    \vspace{-0.3in}
    \caption{Example workflow of \sys.}
    \vspace{-0.1in}
    \label{fig:example}
\end{figure*}

%%%%%%%%%%%%%%%%%%%%%%%%%%%%%%%%%%%%%%%%%%%%%%%%%%%%%%%%%%%%%%%%%%%%%%%%%%%%%%%%%%%%%%%%%%%%%%%%%%%

\subsubsection{Example Workflow}
\label{sec:system-example}

Figure~5 shows how CacheSolidarity handles prefix reuse across three users issuing requests at timestamps $t_1$–$t_4$. Prompts differ only in private information (e.g., names). Circles denote cache entries and arrows denote prefix sequences; red, green, and orange indicate misses, hits, and flagged entries, respectively. We describe the execution at each timestamp next:

\begin{enumerate}[leftmargin=1.2em]
  \item[\textbf{$t_1$:}] User~1 submits a prompt, creating new cache entries tagged with \texttt{OwnerID = 1}. All entries are unflagged.
  
  \item[\textbf{$t_2$:}] User~2 reuses the first two entries. \sys detects a different \texttt{OwnerID} and flags the second entry. New entries are created beyond this point for User~2.
  
  \item[\textbf{$t_3$:}] User~1 issues a follow-up request. As the owner, reuse continues beyond the flagged entry without restriction.
  
  \item[\textbf{$t_4$:}] User~3 attempts to reuse the prompt with modified private data. Since the prefix is flagged and owned by another user, reuse stops and new entries are created from that point.
\end{enumerate}

%In summary, this example illustrates \sys{}’s selective isolation: flagged cache entries restrict cross-user continuation, diverging into new secure paths for non-owners, while allowing full reuse along the original path for the owner.
%
%We stress that \sys{} always allows owners to extend their prefixes even though some nodes might be flagged to ensure that benign requests are not delayed.
%
%For instance, User~2 would be allowed to further extend their own path 1-2-6-7-8 although node 2 is flagged.
In summary, this example illustrates \sys{}’s selective isolation: flagged prefixes prevent cross-user continuation, forcing non-owners onto separate paths while allowing full reuse for the owner. For example, User~2 can still extend its own path (1-2-6-7-8) even if node~2 is flagged.
% 

%%%%%%%%%%%%%%%%%%%%%%%%%%%%%%%%%%%%%%%%%%%%%%%%%%%%%%%%%%%%%%%%%%%%%%%%%%%%%%%%%%%%%%%%%%%%%%%%%%%
\subsection{Activator Optimization}
\label{sec:system-load-monitor}

%As shown in Section~\ref{sec:motivation:leaks}, timing side channels due to APC are \emph{conditionally} distinguishable and exploitable:latency gaps between cache hits and misses grow with (i) longer shared prefixes and (ii) larger model sizes, but are (iii) progressively masked as request-per-second (RPS) increases due to batching, queuing, and GPU saturation effects, (iv) while the hardware platform directly influences timing differences. Therefore, a defense that is always-on would over-protect under high load and impose unnecessary overhead on benign users.
As shown in Section~\ref{sec:motivation:leaks}, APC timing side channels are \emph{conditionally} exploitable: latency gaps between cache hits and misses increase with longer prefixes and larger models, but diminish under higher system load, while also depending on the underlying hardware. As a result, ``always-on'' defenses (Table~\ref{tab:related-works}) may incur unnecessary overhead.
%\sys addresses this by allowing system administrator to conditionally \emph{activate} selective isolation only when timing signals are distinguishable for the current operating point (model, hardware, workload). To achieve this, \sys continuously monitors the latency gap between hits and misses over a sliding time window and computes the \emph{Kernel Density Estimation (KDE) overlap} between their distributions. The KDE overlap~\cite{kde-overlap} is a statistical measure of similarity that indicates how distinguishable two distributions are. Values close to 100\% mean the two distributions are nearly identical (i.e., hits and misses are statistically indistinguishable), while values near 0\% mean they are well separated (i.e., an attacker can easily tell them apart).
\sys addresses this by activating selective isolation only when timing differences are distinguishable under the current operating conditions (model, hardware, workload). %To do so, it continuously measures the latency gap between hits and misses over a sliding window of past history and computes the \emph{KDE overlap}, a statistical measure of how distinguishable the two distributions are~\cite{kde-overlap}. High overlap (values close to 100\%) indicate indistinguishable behavior, while low overlap (values close to 0\%) indicates exploitable leakage (i.e., an attacker can easily tell apart the timing differences).
{To achieve this, it relies on an offline profiling phase combined with lightweight online monitoring.}

{
Offline, \sys profiles the underlying hardware–LLM pair by running workloads with varying request rates and prompt lengths, measuring latency distributions for cache hits and misses, and computing the \emph{KDE overlap}, a statistical measure of how distinguishable these distributions are~\cite{kde-overlap}. It then stores a mapping $(\text{RPS}, \text{prompt length}) \rightarrow \text{overlap}$, where inputs are bucketed as shown in Figure~\ref{fig:side-channels}: RPS in increments of 1, and prompt lengths in increments of 50.}

{Online, \sys maintains a sliding window $h$ of recent requests and calculates the request rate and average prompt length, given that the underlying the hardware and LLM are fixed. Then it uses the profiled mappings $(\text{RPS}, \text{prompt length}) \rightarrow \text{overlap}$ to retrieve the KDE overlap($h$). If the current $(\text{RPS}, \text{prompt length})$ pair has not been profiled, \sys defaults to enabling security and skips the threshold comparison described next.}

The Activator is configured with a static threshold $\theta$. When the KDE overlap($h$) exceeds $\theta$, \sys disables selective isolation and allows full reuse, while still updating cache metadata. When the overlap falls below $\theta$, \sys activates detection and prefix isolation. This decision is updated per request using a sliding window of TTFT samples, enabling dynamic adaptation to workload conditions. The choice of $\theta$ is left to the administrator and reflects the security–performance trade-off: higher thresholds (up to 100\%) increase security but may reduce performance.

{Section~\ref{sec:eval:sensitivity} evaluates the impact of the threshold selection to the security and performance of the system and provides guidelines to the system administrators. Section~\ref{sec:security_activator} discusses how the Activator's design mitigates potential adversarial threshold manipulation.}

%%%%%%%%%%%%%%%%%%%%%%%%%%%%%%%%%%%%%%%%%%%%%%%%%%%%%%%%%%%%%%%%%%%%%%%%%%%%%%%%%%%%%%%%%%%%%%%%%%%

\section{Security Analysis}
\label{sec:system-sec-analysis}

\newcommand{\cachestate}{\mathit{c}}
\newcommand{\atk}{\mathit{atk}}
\newcommand{\hit}{\mathit{H}}
\newcommand{\miss}{\mathit{M}}
\newcommand{\request}{\mathit{r}}
\newcommand{\obs}{\mathit{obs}}

In this section, we analyze the security guarantees provided by \sys{}. First, we precisely characterize its selective isolation guarantees, and then examine the implications of the activator.

\subsection{Security of Selective Isolation}
\label{sec:security_isolation}

{
We characterize which KV cache states an attacker can distinguish via APC timing leaks under \sys{}’s prefix-based selective isolation.
\subsubsection*{Cache states}
We consider a simplified model of the KV cache state, extended with \sys{}’s metadata.
A cache state $\cachestate$ is a set of cache entries, each one consisting of a tuple $(\mathit{id}$, $\mathit{data}$, $\mathit{ownerId}, \mathit{attackFlag}$, $\mathit{next})$. % as described in \Cref{sec:system-detector}.
Two states $\cachestate_1, \cachestate_2$ are \emph{structure-equivalent for user $u$} if they are identical except for the $\mathit{data}$ associated with entries not owned by $u$.

\subsubsection*{Attacker model}

We consider an attacker $\atk$ who interacts with the system as a regular user and issues requests alongside other users.
For an issued request $\request$, the attacker $\atk$ can observe whether the prefixes associated with the request result in a sequence of \emph{hits}~($\hit$) and/or misses~($\miss$). The sequence may consist of only hits $\hit$...$\hit$, only misses $\miss$...$\miss$, or a sequence of hits followed by misses $\hit\hit$...$\miss\miss$ (some prefixes hit the cache). 
%
%That is, the attacker has access to a perfect APC timing channel.
%
Given the current state $\cachestate$ and the request $\request$ issued by $\atk$, we define $\obs(\cachestate,\atk,\request) \in \{ \hit,\miss\}^*$ to be an attacker observation (i.e., the sequence of $\hit$ and $\miss$).
\subsubsection*{Security guarantees}
To characterize \sys{}'s guarantees, we first need to introduce two properties of requests. First, a request $\request$ results in a \emph{first-entry hit}, defined as $\mathit{firstHit}(\cachestate,\atk,\request)$, if there is a cache entry for the first prefix in the request $\request$. This results in a hit in state $\cachestate$, i.e., the first element of $\obs(\cachestate,\atk,\request)$ is $\hit$. Second, a request $\request$ results in an \emph{unflagged hit}, written $\mathit{unflagHit}(\cachestate,\atk,\request)$, if it results in a hit on an unflagged entry in state $\cachestate$, i.e., if $\obs(\cachestate,\atk,\request)$ contains an $\hit$ corresponding to an unflagged cache entry. We now characterize the security guarantees of \sys{}’s selective isolation by identifying which pairs of states an attacker can distinguish. This characterization is expressed as a non-interference property, a standard way to formalize confidentiality~\cite{goguen1982security,goguen1984unwinding,denning1976lattice,volpano1996sound,sabelfeld2003language}.
\begin{theorem}\label{theorem:sec-guarantees}
Let $\request$ be an $\atk$-request and $\cachestate_1, \cachestate_2$ be  cache states.
If:
\begin{enumerate}[topsep=0pt,itemsep=-1ex,partopsep=1ex, parsep=1ex]
\item $\cachestate_1$ and $\cachestate_2$ are structure-equivalent for $\atk$,
\item $\neg \mathit{firstHit}(\cachestate_1,\atk,\request) \wedge \neg \mathit{firstHit}(\cachestate_2,\atk,\request)$, and
\item  $\neg \mathit{unflagHit}(\cachestate_1,\atk,\request) \wedge \neg \mathit{unflagHit}(\cachestate_2,\atk,\request)$,
\end{enumerate}
%(1) $\cachestate_1$ and $\cachestate_2$ are structure-equivalent for $\atk$, (2) $\neg \mathit{firstHit}(\cachestate_1,\atk,\request) \wedge \neg \mathit{firstHit}(\cachestate_2,\atk,\request)$, and (3)  $\neg \mathit{unflagMatch}(\cachestate_1,\atk,\request) \wedge \neg \mathit{unflagMatch}(\cachestate_2,\atk,\request)$, 
\quad then $\obs(\cachestate_1,\atk,\request) = \obs(\cachestate_2,\atk,\request)$.
\end{theorem}

\noindent Theorem~\ref{theorem:sec-guarantees} states that, for all requests  satisfying conditions (2) and (3), \sys{} prevents attackers from inferring any information about data stored in cache entries not owned by them.} These guarantees are conditional by design, reflecting its lightweight system design and security/performance trade-off.
%through APC timing channels induced by APC hits/misses.

Next, we describe the two cases not covered by \sys{}:
{
\begin{tightitemize}
  \item \textit{Attacks targeting the first entry (condition (2)):} 
  \sys{} does not protect against attacks targeting the first cache entry of a request $r$ that result in a $\mathit{firstHit}$. %i.e., against cases where there is no (non-empty) prefix. 
  This happens because in this case there is no parent node (Figure~\ref{fig:example}) that \sys{} can flag as potentially involved in an attack. Introducing a ``dummy'' parent node would make \sys{} fall back to the naive user-level cache isolation.
  \item \textit{Attacks against unflagged prefixes (condition (3)):} 
  \sys{} does not protect \emph{unflagged} cache entries that result in an $\mathit{unflagHit}$, i.e., entries that \sys{} has not yet flagged because their first reuse occurs by the attacker, who correctly guesses the prefix, resulting in a cache hit. This is the only case where an $\mathit{unflagHit}$ can happen, because \sys conservatively flags and isolates any prefixes exactly after they are reused by different users.
  The only way to prevent this case is to always treat a cache entry as flagged. However, this would make \sys{} fall back to the naive user-level cache isolation.%, which introduces too high overhead.
\end{tightitemize}
}

\noindent Beyond the two cases described above, \sys{} secures the system against APC leaks. Its guarantees do not rely on semantic detection (as in~\cite{chu2026selectivekvcachesharingmitigate}), but on monitoring prefix overlap across users. Moreover, these guarantees hold regardless of whether attacks are performed by a single or multiple attackers, interleaved with benign requests, or under different initial cache states.

\begin{figure}[t]
\centering

\begin{subfigure}{0.48\linewidth}
    \centering
    \includegraphics[width=\linewidth]{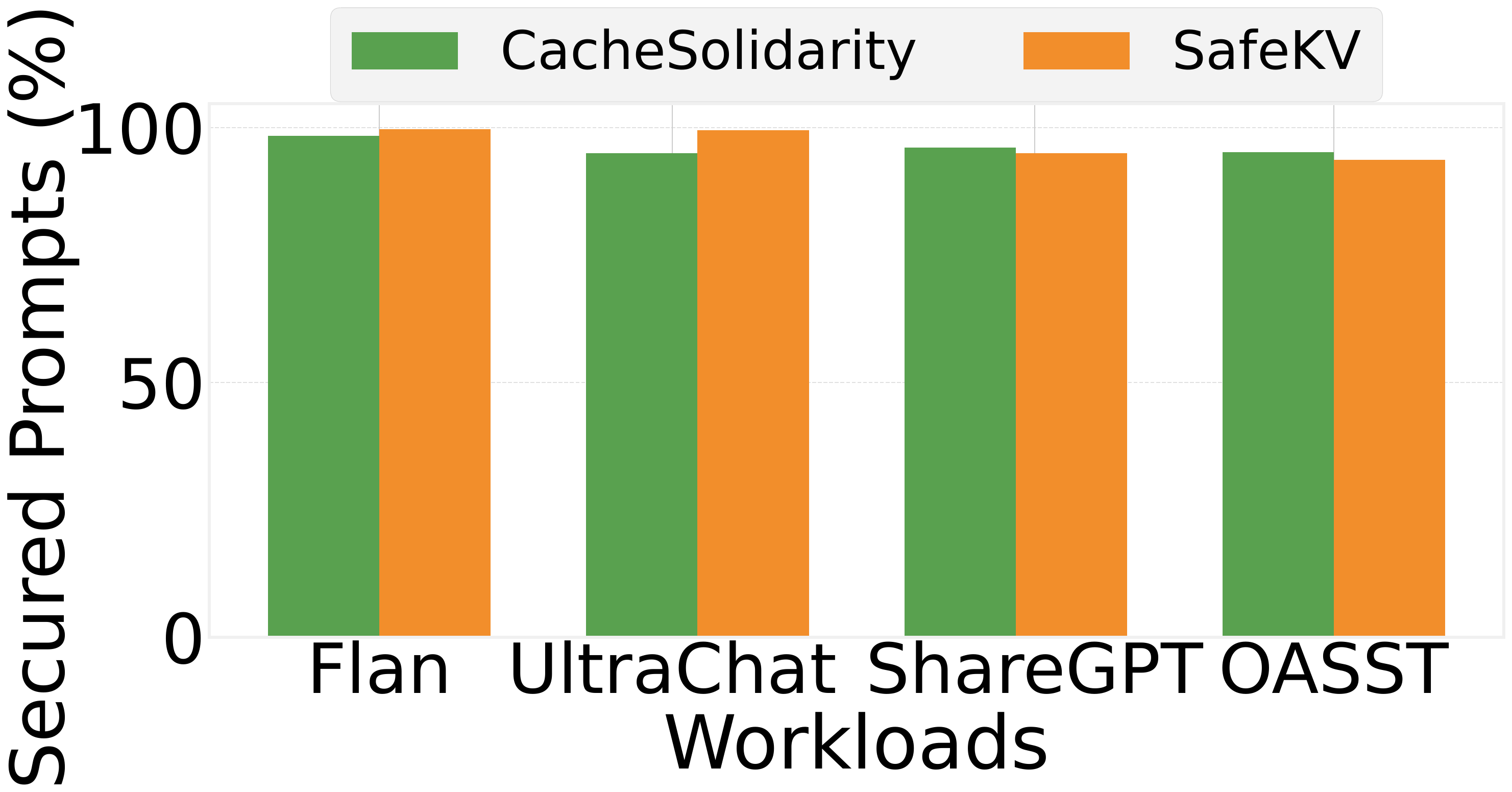}
     \caption{Secured prompts (\%) for \sys vs.\ SafeKV.}
    \label{fig:security_plot}
\end{subfigure}
\hfill
\begin{subfigure}{0.48\linewidth}
    \centering
    \small
    \begin{tabular}{|l|c|c|}
    \hline
    \textbf{Dataset} & \textbf{1st Entry} & \textbf{1st Try} \\
    \hline
    Flan & 0.12\% & 1.51\%  \\
    UltraChat & 0.01\% & 5.03\% \\
    ShareGPT & 1.35\% & 2.63\% \\
    OASST1 & 1.08\% & 3.8\% \\
    \hline
    \end{tabular}
     \caption{Failure cases for \sys (first-entry, first-try).}
    \label{fig:security_table}
\end{subfigure}
\vspace{-10pt}
% \caption{{When $\theta = 100\%$ and the timing side channel is exploitable, \sys only misses first-entry, first-try cases (right), whereas SafeKV fails to secure prompts due to semantic misclassification (left).}}
\caption{Security vs.\ SafeKV ($\theta = 100\%$) and \sys failure cases.}
\label{fig:security_datasets}
\vspace{-0.2in}
\end{figure}

\compactpara{{Occurrence of \sys's limiting conditions.}} {In practice, these conditions are rarely met. We revisit the real datasets in Table~\ref{tab:secret-datasets}, applying the same methodology to identify sensitive content and simulate prompt stealing attacks (Section~\ref{sec:motivation:attacks}). Table~\ref{fig:security_table} reports the fraction of prompts with secrets in the first cache entry (condition (2)) and those recoverable on the first attempt (condition (3)). Only 0.01\%--1.35\% of prompts contain secrets in the first entry. Moreover, short prompts ($<100$ tokens), which are more likely to contain secrets early on, exhibit high KDE overlap (Figure~\ref{fig:KDE_models}), making cache hits and misses harder to distinguish. In practice, prefix caching is typically restricted to long prompts (e.g., $>1024$ tokens in OpenAI~\cite{openai2024promptcaching}), further reducing this risk. For first-try attacks, only 3.24\% of prompts are recoverable, on average. Notably, this {\it assumes an exploitable timing channel}; in practice, the factors that impact the exploitability of the timing side channel discussed in Section~\ref{sec:motivation:leaks}, may further limit attack success. In addition, {\it user solidarity} in \sys{} can further reduce first-try success: if another user reuses the target prefix before the attacker, the prefix gets isolated and is inaccessible to the attacker.
}

\subsection{Security impact of threshold activation}
\label{sec:security_activator}

The Activator disables \sys{}’s selective isolation when the KDE overlap over a sliding window $h$ exceeds a threshold $\theta$, indicating that timing differences are indistinguishable.
This can impact the security of the system in two ways.
First, a low threshold $\theta$ may leave the system unprotected when timing leakage is exploitable.
Second, an attacker may manipulate the Activator by shaping traffic patterns. 
For example, issuing bursts of requests to inflate the KDE overlap and force deactivation, or throttling requests to reduce overlap and trigger isolation for probing.

\sys mitigates these risks through three design choices. First, the Activator computes the KDE overlap over a sliding window, requiring sustained conditions rather than reacting to transient fluctuations. Second, activation decisions rely on aggregate system-level measurements rather than per-user metrics or prompt semantics, making it difficult for an attacker to influence the observed signal. Finally, cache metadata is always updated: shared prefixes remain flagged even when isolation is disabled, preventing attackers from manipulating cache state.

{While dynamic threshold adaptation is possible, it introduces a distinct risk since attackers might shape traffic patterns over time (e.g., sustained high or low load) to steer the system into adapting its threshold to a specific value. Unlike fixed-threshold designs, this creates a feedback loop where adversarial inputs may influence future system decisions, potentially causing the system to keep isolation deactivated. Even though the sliding window mitigates short-term manipulation, it does not prevent this form of threshold miscalibration. We therefore adopted a static threshold to avoid such feedback-based attacks and ensure predictable behavior, leaving its selection to the administrator following common practice in real-world systems (e.g., Google’s differential privacy deployments~\cite{google_dp}) and the proposed guidelines in Section~\ref{sec:eval:sensitivity}.}
\section{Evaluation}
\label{sec:evaluation}

\begin{figure*}[t]
\centering
\begin{subfigure}[t]{0.32\textwidth}
    \centering
    \includegraphics[width=\linewidth]{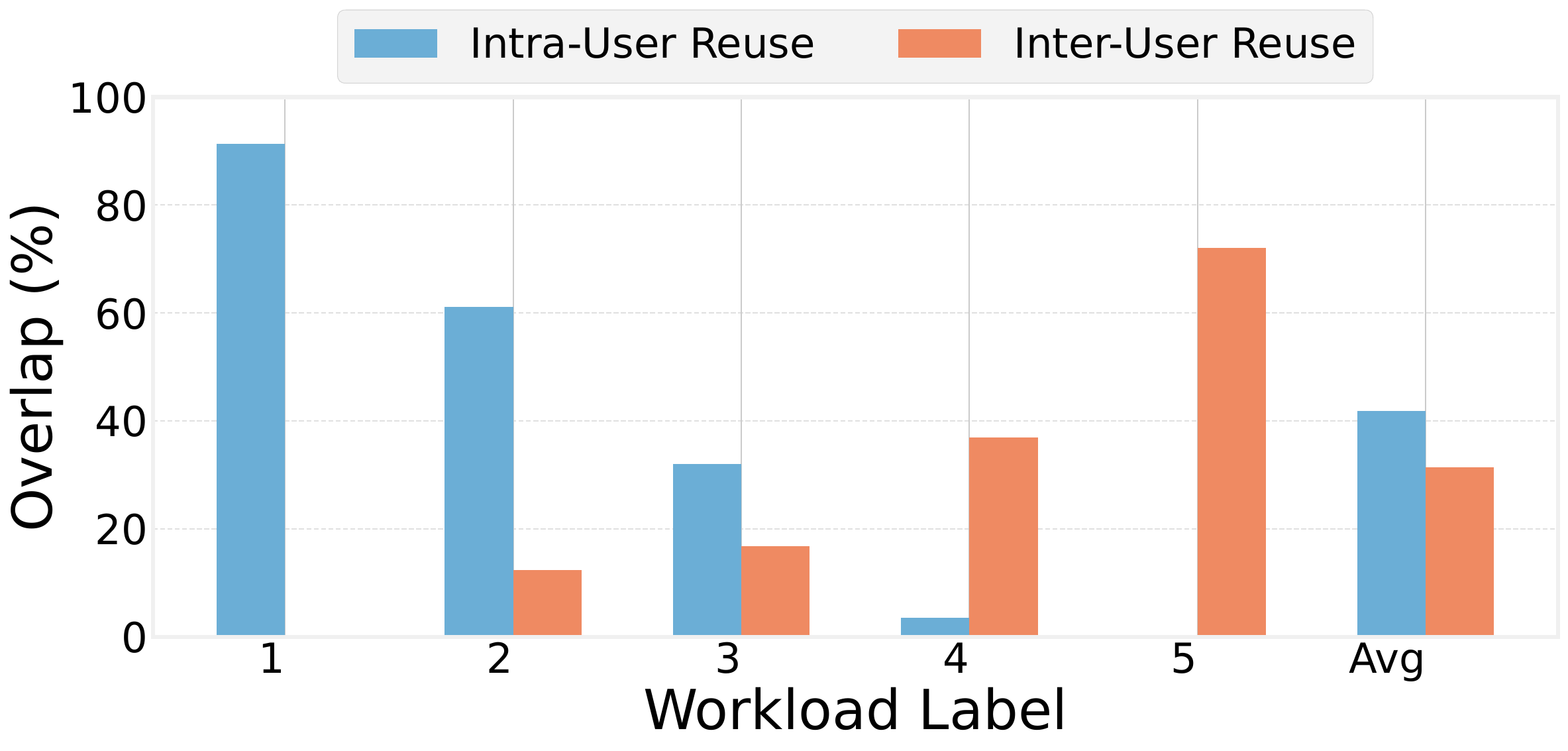}
    \caption{Workload construction.}
    \label{fig:overlap}
\end{subfigure}\hfill
\begin{subfigure}[t]{0.32\textwidth}
    \centering
    \includegraphics[width=\linewidth]{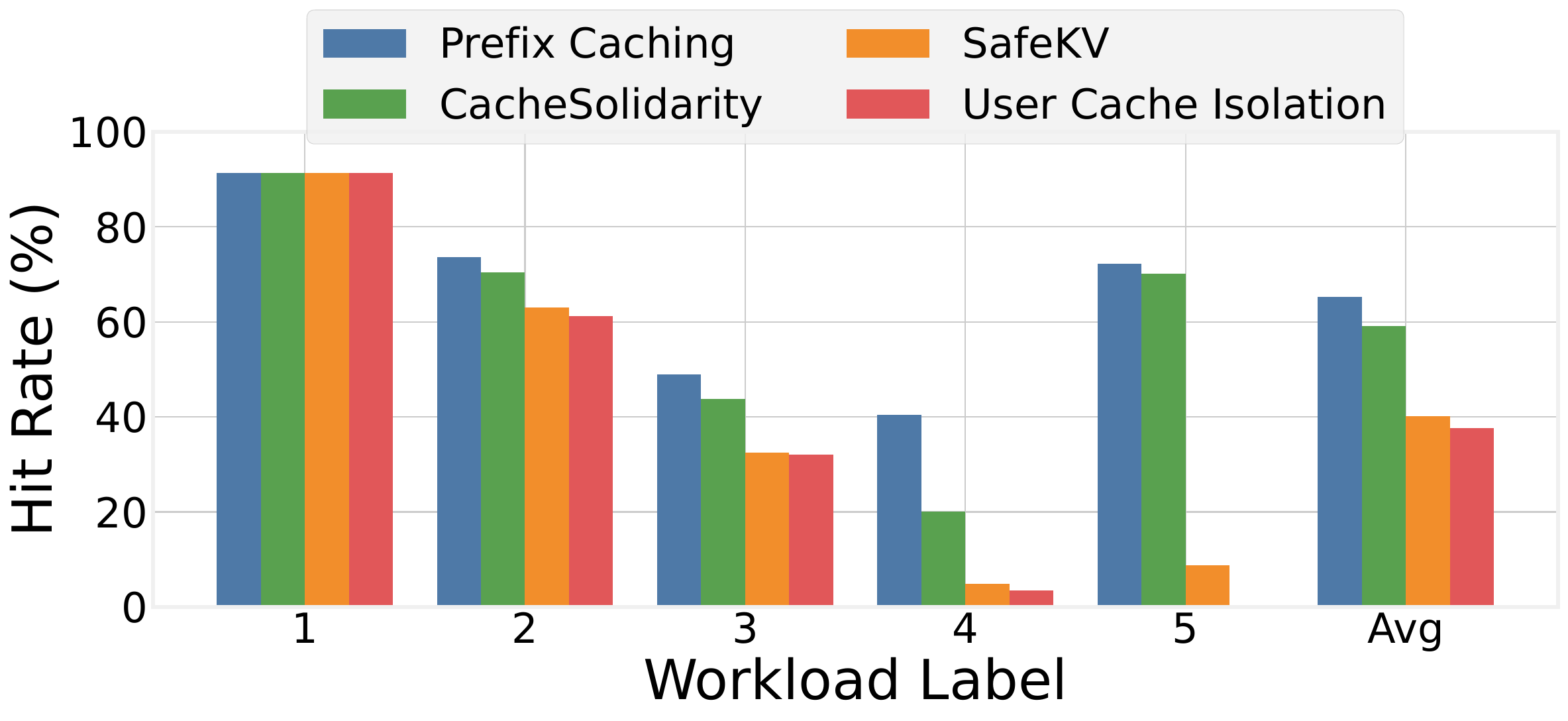}
    \caption{Cache hit rate.}
    \label{fig:hit_rate}
\end{subfigure}
\begin{subfigure}[t]{0.32\textwidth}
    \centering
    \includegraphics[width=\linewidth]{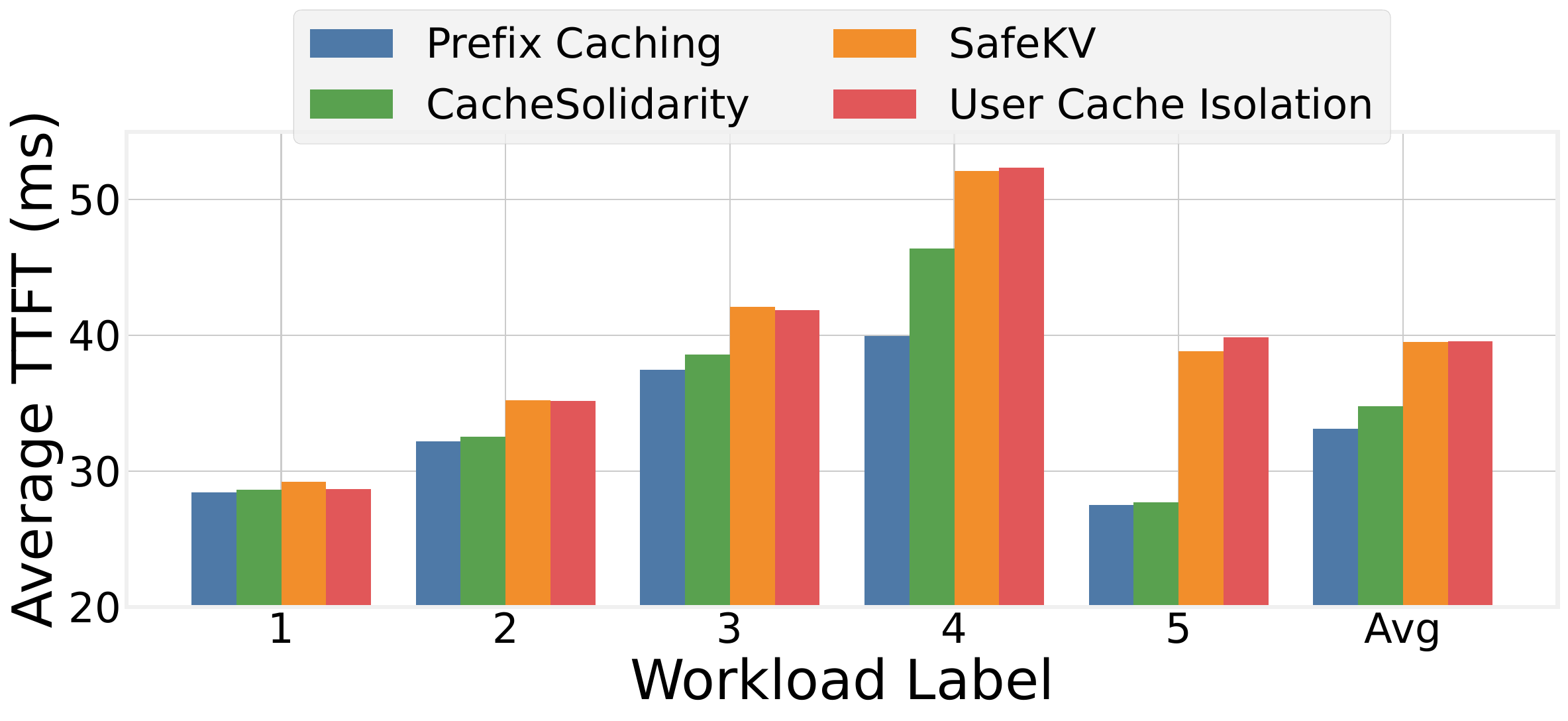}
    \caption{Average TTFT.}
    \label{fig:ttft}
\end{subfigure}\hfill
\vspace{-10pt}
\caption{{Workload construction and inference performance comparison across workloads and baselines}.}
\label{fig:eval_workloads}
\end{figure*}

We evaluate \sys to understand its impact on performance and security in multi-tenant LLM serving environments. Our goal is to quantify the performance benefits under realistic workloads, across LLM families and sizes, while examining the overhead introduced by \sys. Specifically, we address the following research questions:
\begin{description}
\item[RQ1:] {\bf Performance Gains:} How much does \sys{} accelerate inference compared to existing defenses across different workloads and models?% (Sections~\ref{sec:eval_perf:across-workloads},~\ref{sec:eval_perf:across-models},~\ref{sec:eval_perf:across-rps})?
\item[RQ2:] {\bf Overheads:} What are the latency and resource overheads introduced by \sys{}?% (Sections~~\ref{sec:eval_perf:across-workloads},  ~\ref{sec:eval_perf:across-models}, and \ref{sec:eval_scale})?
\item[RQ3:] {\bf Security:} Can \sys{} effectively secure prompts that contain private information?% (\ref{sec:eval:attacker}) 
\item[RQ4:] {\bf Sensitivity:} What is the impact of the activation threshold on \sys{}'s performance {and security}? %(\ref{sec:eval_sensitivity})
\end{description}

%%%%%%%%%%%%%%%%%%%%%%%%%%%%%%%%%%%%%%%%%%%%%%%%%%%%%%%%%%%%%%%%%%%%%%%%%%%%%%%%%%%%%%%%%%%%%%%%%%%
\subsection{Experimental Setup}

Table~\ref{tab:models} summarizes the LLM models used for evaluation, their memory footprint, and the remaining GPU memory available for hosting the KV cache. The serving system is vLLM~\cite{vllm} (version 0.8.5) with a block size of 16. We evaluate the performance of \sys on a native hardware server equipped with an NVIDIA A100 GPU (40 GB memory). {For larger models, we use a server with 4 H100 GPUs (94 GB each). We evaluate Llama3-70B that needs 131.5 GB of memory, so we configure it with tensor parallelism TP = 2.}

\compactpara{Workloads.} 
For our experiments, we use real prompts from the ShareGPT~\cite{anon8231489123_2023} dataset. We process prompts using the state-of-the-art BERT model~\cite{devlin2019bert} (google-bert/bert-base-uncased) to (i) identify and mask sensitive words that contain private information, and (ii) replace masked words with alternative candidates. This allows us to evaluate prompts that share common prefixes but differ in private content (e.g., names, addresses etc).

Then, we construct different workloads by sampling prompts, varying the level of prefix overlap. Higher overlap means a longer shared prefix, which leads to more cache reuse and better performance. For example, private data at the end of the prompt (``You are a helpful assistant. I want you to write an email to reply to [sensitive information]'') creates high overlap, while at the beginning (``My name is [sensitive information]'') results in minimal overlap. To capture a multi-user environment, we vary both the \textit{intra-user} reuse (how often a user reuses a previously cached prefix) and \textit{inter-user} reuse (how much different users share common prefixes). Figure~\ref{fig:overlap} illustrates these dimensions across our constructed workloads. 

For the experiments in Section~\ref{sec:eval_perf}, each workload includes 10 users issuing 100 requests each (1000 total), with arrivals following a Poisson process at a rate of 10 requests per second (RPS = 10), reflecting realistic LLM workloads~\cite{kwon2023pagedattention, sarathi, dist-serve, splitwise}. The Activator of \sys is configured with $\theta = 100\%$ and window $h = 1$\,second. One of the users acts as an adversary targeting a common prompt template, issuing variants with different candidate values at the position of the secret (as described in Section~\ref{sec:motivation:attacks}).

\compactpara{Evaluation Baselines.} We compare \sys against the following baselines, based on the categorization in Table~\ref{tab:related-works}:

\begin{tightitemize}
    \item {\bf Automatic Prefix Caching (APC):} Enables reuse of common prefixes cached across all users~\cite{vllm_prefix_caching}. This approach maximizes reuse and efficiency but offers no isolation or security guarantees. 
    \item {\bf User Cache Isolation:} Allocates dedicated KV cache regions per user to ensure isolation and security~\cite{cache-partitioning, input-snatch, pmlr-v267-gu25b}. This eliminates the benefits of caching common prefixes, reducing efficiency.
    \item {{\bf Selective Sharing:} Uses semantic analysis (rule-based checks and LLM-based validation) to classify prompts as sensitive and selectively isolate them, sharing only non-sensitive ones, as proposed in SafeKV~\cite{chu2026selectivekvcachesharingmitigate}.}
    \item {\bf \sys:} Our proposed mechanism that isolates prefixes based on sharing behavior across users, without relying on semantics, balancing security and performance.
\end{tightitemize}

\vspace{-0.055in}

\compactpara{Implementation Details of the Baselines.}
\sys is implemented on top of vLLM~\cite{vllm} (version 0.8.5) with the V1 engine for multi-tenancy. % and privacy-aware caching.
To extend the KV cache as described in \ref{sec:system-detector}, we extend the KVCacheBlock class with the two metadata fields.
% We also extend in the KV cache and allowing to communicate with the Detector.
We introduce a new Detector component, as part of the Scheduler that detects malicious requests before scheduling and updates the new metadata fields \texttt{attack\_flag} and \texttt{owner\_id}.
We extend the metrics collection of vLLM to support the Activator component which is implemented in the asynchronous LLM execution layer. %\sys is extensively documented and will be open-sourced to encourage community adoption and future extensions. 

User cache isolation is implemented by modifying vLLM to prepend a unique user identifier (a single token) to each request, effectively preventing prefix sharing across users.

{The semantic-aware selective sharing baseline, SafeKV~\cite{chu2026selectivekvcachesharingmitigate}, is built on SGLang’s tree-based architecture~\cite{sglang} and is not open-sourced. We therefore implement it on top of vLLM~\cite{vllm} (version 0.8.5) V1 engine, following the methodology described in the paper, to the best of our understanding.}

{SafeKV consists of three components. First, an asynchronous three-stage detection pipeline that decides whether each KV entry is private or shareable. It uses (i) a rule-based detector covering 34 categories of sensitive information (pii-extract 0.0.2~\cite{villegas_pii_extract}), (ii) a lightweight LLM-based classifier (Llama‑3.2‑1B~\cite{llama32024}) for cases not matched by these rules, and (iii) a final step that uses the serving LLM to resolve remaining uncertain cases. Each stage performs a pass over the prompt to identify sensitive content and, when found, it produces a structured output (e.g., a JSON record specifying whether private information is present, along with its position and semantic category). Second, a cache index that enforces access control using per-entry privacy and ownership metadata, similar to \sys. Third, a runtime safeguard that monitors user access patterns, specifically whether entries are repeatedly accessed by only a few users, and disables sharing when such behavior appears suspicious. All new cache entries are marked private by default, while the detection pipeline runs asynchronously. Once it completes, only the non-sensitive prefix will be marked as shareable. The runtime safeguard continuously runs in the background. Finally, while SGLang enforces metadata at token granularity, the vLLM implementation operates at the KV block level. }

%%%%%%%%%%%%%%%%%%%%%%%%%%%%%%%%%%%%%%%%%%%%%%%%%%%%%%%%%%%%%%%%%%%%%%%%%%%%%%%%%%%%%%%%%%%%%%%%%%%
\subsection{{Comparison Against Baselines}}
\label{sec:eval_perf}

In this section, we compare \sys against the baselines, focusing on inference performance across workloads and models, system overheads, and security.

\subsubsection{Across Workloads}\label{sec:eval_perf:across-workloads}

Figure~\ref{fig:eval_workloads} reports average time-to-first-token (TTFT) and cache hit rate across the five representative workloads (Figure~\ref{fig:overlap}) and their aggregate average. The reported values exclude the attacker. This experiment runs inference of Llama-13B on a single A100 GPU. {Hence, for the SafeKV baseline, the validation LLM runs on the same GPU as the inference model, competing for GPU cycles and reducing the effective KV cache capacity compared to \sys (Table~\ref{tab:models}). In addition, prefix reuse across users is delayed until the asynchronous semantic validation completes (either rule- or LLM-based), further limiting potential cache hits. Across workloads, most secrets (62\%--81.3\%) are detected by the rule-based Tier-1, with the remainder handled by Tier-2 (the validation LLM, Llama-3.2-1B). As inter-user reuse increases, more repeated prompts are handled by Tier-1, reducing reliance on LLM-based validation. For the average workload, 63.4\% of secrets are detected by Tier-1 and 36.6\% by Tier-2, with average latencies of 8.1 ms and 2.6 sec,  respectively; Tier-3 is never reached. Although Tier-2 uses a lightweight LLM, its latency is dominated by response generation (i.e., JSON output) and by GPU contention, as it runs alongside the inference model on the same GPU. These semantic validation delays alongside the reduced KV cache capacity, combined with LRU eviction, may cause less frequently reused prefixes to be evicted before they are validated and shared across users.\looseness=-1}

\compactpara{\bf Workloads with high intra-user reuse.} %Starting with one of the extreme cases, 
In {\it Workload~1} %exhibits high intra-user reuse and zero inter-user reuse, meaning that each 
where users repeat similar prompts but do not share common prefixes, %with other users. As expected, for this workload 
all baselines perform similar to cache isolation. %{SafeKV shows slightly higher TTFT due to resource contention from semantic validation.}
Similarly, {\it Workload~2} %maintains high intra-user reuse but 
introduces moderate inter-user reuse. As expected, User Cache Isolation suffers lower hit rate because it misses shared prefixes, while \sys stays very close to Prefix Caching, preserving most performance benefits. {SafeKV has slightly reduced hit rate, thus slightly higher TTFT.}

\compactpara{\bf Workloads with high inter-user reuse.} {In {\it Workload~4}, where inter-user reuse dominates but some intra-user reuse remains, both \sys and SafeKV experience reduced hit rates. However, SafeKV incurs higher TTFT, revealing a \textit{key limitation of semantic-based isolation} beyond resource contention: prefixes under semantic validation cannot be reused across users, causing them to miss potential hits and, under high reuse, be evicted before reuse even occurs. This issue does not arise in \sys, which enables immediate prefix reuse.} 
{\it Workload~5} has high inter-user reuse and zero intra-user reuse, meaning that users ask similar prompts to each other but not to themselves, leading to zero hit rate of cache isolation. %This explains why User Cache Isolation has zero cache hit rate, leading to cache misses and the highest TTFT among all baselines. Most importantly, this workload {\bf stresses the critical path} of \sys by activating the detector on every request due to the high inter-user reuse. 
In addition, we see that \sys's lightweight design does not hurt performance: TTFT remains very close to Prefix Caching, and hit rate is similarly high. %, demonstrating that \sys can enforce security without sacrificing efficiency even under %worst-case conditions.
{SafeKV, for the reasons mentioned already, shows reduced performance.}

\compactpara{\bf Average Case.} 
 Across all workloads, \sys significantly improves over User Cache Isolation while remaining within 5--10\% of Prefix Caching, {whereas SafeKV consistently incurs lower hit rates and higher TTFT due to the delay in reuse until semantic validation finishes and the reduced cache capacity in a single GPU.} The hit-rate gap to Prefix Caching reflects {\it false positives}: safe prefixes isolated because CacheSolidarity relies on reuse behavior, not semantics.

%%%%%%%%%%%%%%%%%%%%%%%%%%%%%%%%%%%%%%%%%%%%%%%%%%%%%%%%%%%%%%%%%%%%%%%%%%%%%%%%%%%%%%%%%%%%%%%%%%%
\subsubsection{Across LLM Models}
\label{sec:eval_perf:across-models}

\begin{table}[t]
\centering
\small
\setlength{\tabcolsep}{3pt}

\begin{tabular}{l|c|cc|cc}
\textbf{Model (Abbrev.)} & 
\textbf{Model} &
\multicolumn{2}{c|}{\textbf{\sys}} &
\multicolumn{2}{c}{\textbf{SafeKV}} \\
&
&
\textbf{GB} & \textbf{\%} &
\textbf{GB} & \textbf{\%} \\

\midrule

Gemma3-4B~\cite{gemma3_4b_it}      & 8.6\,GB  & 31.4\,GB & 78.5\% & 28.8\,GB & 77.0\% \\
Llava-0.5B~\cite{llava_onevision_qwen2_0_5b_ov_hf} & 1.68\,GB & 38.32\,GB & 95.8\% & 35.72\,GB & 95.5\% \\
Llava-7B~\cite{llava_onevision_qwen2_7b_ov_chat_hf} & 15.03\,GB & 24.97\,GB & 62.4\% & 22.37\,GB & 59.8\% \\
Llama2-7B~\cite{llama2_7b_chat_hf} & 12.6\,GB & 27.4\,GB & 68.5\% & 24.8\,GB & 66.3\% \\
Llama2-13B~\cite{llama2_13b_chat_hf} & 24.3\,GB & 15.7\,GB & 39.3\% & 13.1GB & 35.0\% \\
{Llama3-70B~\cite{meta2024llama31}} & {131.5 GB} & {56.6 GB} & {30.05\%} & {56.5 GB} & {30.05\%}\\
Qwen2-2B~\cite{qwen2_vl_2b_instruct} & 4.15\,GB & 35.85\,GB & 89.6\% & 33.25\,GB & 88.9\% \\
Qwen2.5-3B~\cite{qwen2_5_vl_3b_instruct} & 7.16\,GB & 32.84\,GB & 82.1\% & 30.24\,GB & 80.9\% \\
Qwen2-7B~\cite{qwen2_vl_7b_instruct} & 15.53\,GB & 24.47\,GB & 61.2\% & 21.87\,GB & 58.5\% \\
Qwen2.5-7B~\cite{qwen2_5_vl_7b_instruct} & 15.63\,GB & 24.37\,GB & 60.9\% & 21.77\,GB & 58.2\% \\

\end{tabular}

\caption{{Evaluated LLM models and memory allocation.}}
\label{tab:models}
\vspace{-0.1in}
\end{table}

\begin{figure}[t]
    \centering
    % \begin{subfigure}[t]{0.48\textwidth}
        \centering
        \includegraphics[width=0.9\linewidth]{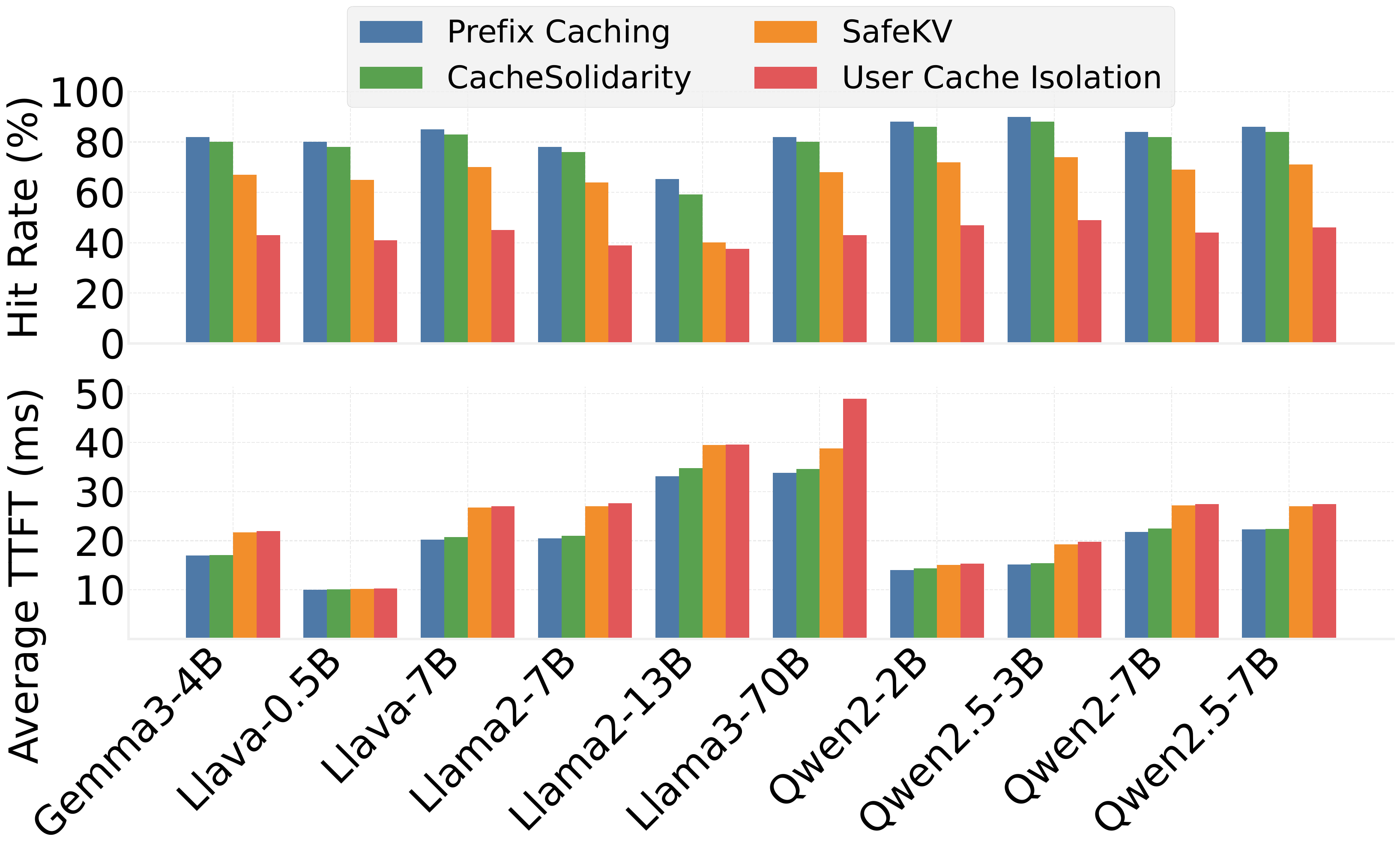}
        \label{fig:eval_models_hit}
        % \caption{Cache hit rate.}
    % \end{subfigure}
    \vspace{-10pt}
     \caption{Hit rate and TTFT across LLMs.}
     \label{fig:eval_models}
    \vspace{-0.2in}
\end{figure}

Figure~\ref{fig:eval_models} reports cache hit rate and TTFT of \sys and the baselines across LLMs (Table~\ref{tab:models}), evaluated on a single A100 GPU under the average workload (Figure~\ref{fig:overlap}). Model size affects both KV cache capacity (larger models leave less space) and inference speed, with smaller models achieving lower TTFT~\cite{konpap-euromlsys24}. 
{For SafeKV, the LLM-based validator runs on the same GPU for all A100 experiments, as done in Section~\ref{sec:eval_perf:across-workloads}. However, for Llama3-70B, which runs on a 4 H100 GPU server with TP = 2, the validator executes on a separate GPU, avoiding resource contention and preserving KV cache capacity.}

%As discussed in Section~\ref{sec:motivation} and shown in Figure~\ref{fig:eval_models_ttft}, 
We observe that \sys effectively closes the performance gap between Prefix Caching and User Cache Isolation  %widens as model size increases and also varies by model family. Across all models, \sys effectively closes this gap, 
remaining within a 6\% TTFT margin of Prefix Caching, while significantly outperforming User Cache Isolation. %This demonstrates that \sys's selective isolation delivers secure and fast inference. 
{Two cases in the performance comparison of the baselines stand out. First, }
%In Figure~\ref{fig:eval_models_hit}, cache hit rates are similar across models except for LLaMA-13B. For this large model, 
for Llama2-13B the KV cache size is very small (Table~\ref{tab:models}), almost half the size compared to the rest of the models, causing frequent evictions and more cache misses, reducing the hit rate of \sys. {In this case, SafeKV exhibits the lowest hit rate due to the combined effect of even smaller effective cache capacity and resource contention with its LLM-based validator running on the same A100 GPU.} 

{However, this limitation is mitigated when executing the Llama3-70B model on a 4×H100 GPU server. This is the only case where SafeKV performs significantly better than cache isolation and closer to \sys. This is because the validator runs on a separate GPU, eliminating resource contention, while the longer inference time of the 70B model hides semantic validation overheads, also removing the validation-delay effects observed in Section~\ref{sec:eval_perf:across-workloads}.}
In conclusion, the performance of \sys remains close to Prefix Caching {and better than semantic-based alternatives} across all LLM model families and sizes.

\begin{figure}[t]
    \centering
    \includegraphics[width=\columnwidth]{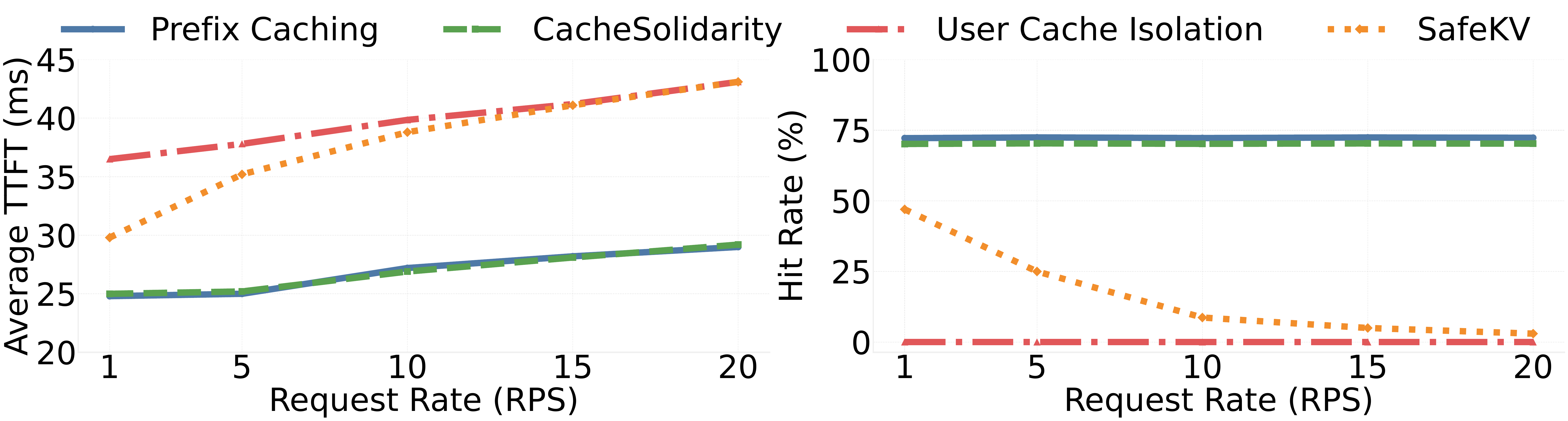}
    \vspace{-20pt}
    \caption{Baseline comparison of TTFT and Hit Rate across increasing RPS for workload 5. }
    \label{fig:eval_baselines_rps}
    \vspace{-0.1in}
\end{figure}

\begin{table}[t]
\centering
\begin{tabular}{lcc}
\toprule
\textbf{Component} & \textbf{Time} & \textbf{Memory} \\
\midrule
Detector & 0.004 ms per request & 32 bytes per cache entry\\
Activator & 0.003 ms & 16 bytes * len(window)\\
%\hline
%Total Overhead & 0.007ms & 0.0248 \\
%\hline
%TTFT & 28.848 & 100.0 \\
\bottomrule
\end{tabular}%
% \caption{Time and memory overheads introduced by the system components of \sys{}.}
\caption{Overhead of \sys{}'s components.}
\label{tab:overhead}
\vspace{-0.3in}
\end{table}

%%%%%%%%%%%%%%%%%%%%%%%%%%%%%%%%%%%%%%%%%%%%%%%%%%%%%%%%%%%%%%%%%%%%%%%%%%%%%%%%%%%%%%%%%%%%%%%%%%%
\subsubsection{{Across Request Rates}}
\label{sec:eval_perf:across-rps}

{
Figure~\ref{fig:eval_baselines_rps} shows the performance comparison across baselines and increasing request rate (RPS) for the Llama2-13B model on a single A100 GPU. We use {\it Workload 5} where any reuse comes from sharing prefixes across users, which directly stresses the mechanisms that enable or delay cross-user reuse. 
Regarding hit rate, it is zero for User Cache Isolation as there is no intra-user reuse, while \sys remains high and close to Prefix Caching. For these baselines, hit rate stays constant as RPS increases, since reuse patterns remain unchanged. In contrast, SafeKV’s hit rate drops with increasing load due to limited KV cache capacity and semantic validation delays: prefixes are blocked from reuse during validation, leading to delayed hits, coupled with higher request rates that lead to higher cache eviction rates under LRU. 
This is reflected in the average TTFT, which increases more sharply for SafeKV compared to the other baselines, where a modest increase is expected due to the higher load.}

 \subsubsection{Overheads.}\label{sec:eval:overheads}
 Table~\ref{tab:overhead} summarizes the operational overheads introduced by the two components of \sys{} in terms of latency and memory resources.
 In terms of per-request latency, \sys{}'s modifications over vanilla vLLM (i.e., the selective isolation scheme and the activator component)  introduce an  average per-request overhead of only 0.007 ms (0.004 ms due to the detector component and 0.003 ms due to the activator component).
 In terms of memory, the metadata introduced by the detector component amount to 32 bytes per cache entry, since each KVCache block was extended with two metadata fields---\texttt{owner\_id} and \texttt{attack\_flag}---of type \texttt{int}.\footnote{Since vLLM is implemented in Python, adding two \texttt{int} fields (as Python objects) to the KVCache class results in an increase of 32 bytes.}
 In contrast, the activator component stores 1 floating point value per request in the sliding window.\footnote{Again, a floating point value in Python takes up to 16 bytes.} 
{Notably, overheads scale with the size of the KV cache (Table~\ref{tab:models}), rather than the number of users, since no per-user state is maintained. SafeKV~\cite{chu2026selectivekvcachesharingmitigate} uses similar metadata to enforce per-user prefix isolation, while Table~\ref{tab:models} also captures its additional storage overhead, when the LLM-based validator (Tier-2) is colocated on the same GPU, reducing the available KV cache.}

\subsubsection{{Security.}} 
\label{sec:eval:security}

{
Figure~\ref{fig:security_plot} compares the fraction of prompts secured by \sys across real-world datasets (Table~\ref{tab:secret-datasets}) with the Activator set to $\theta = 100\%$, against the semantic-aware baseline SafeKV~\cite{chu2026selectivekvcachesharingmitigate}. Overall, \sys achieves high protection (96.12\%), closely matching SafeKV (96.95\%). For SafeKV, across datasets, 62.5\% of prompts are handled by Tier-1 (6.5\,ms average latency), 32.19\% by Tier-2 (Llama-3.2-1B, 1.8\,s), and 2.26\% by Tier-3 (Llama2-13B, 4.5\,s). The higher latency of Tier-2 and Tier-3 is due to LLM-based processing, which scales with the LLM size and includes the time it takes to generate the structured JSON output. The remaining 3.05\% of prompts contain secrets that are not detected due to semantic limitations. SafeKV fails to secure this small subset of prompts due to misclassification, particularly for implicit or context-dependent information (e.g., ``I work in the only bakery in the city''), obfuscated forms (e.g., ``john dot smith at protonmail dot com''), and cross-alphabet, multilingual, or inferred sensitive information (e.g., ``I pick up insulin weekly''). In contrast, \sys successfully protects these cases, but fails in a small subset of first-try, first-entry cases, discussed in Section~\ref{sec:security_isolation}.
}

%%%%%%%%%%%%%%%%%%%%%%%%%%%%%%%%%%%%%%%%%%%%%%%%%%%%%%%%%%%%%%%%%%%%%%%%%%%%%%%%%%%%%%%%%%%%%%%%%%%%%%%%%%%%%%%%%%%%%%%%%%%%%%%%%%%%
\subsection{Sensitivity Analysis}
\label{sec:eval:sensitivity}

\begin{figure}[t]
\centering

\begin{subfigure}[t]{1.0\columnwidth}
    \centering
    \includegraphics[width=0.8\columnwidth]{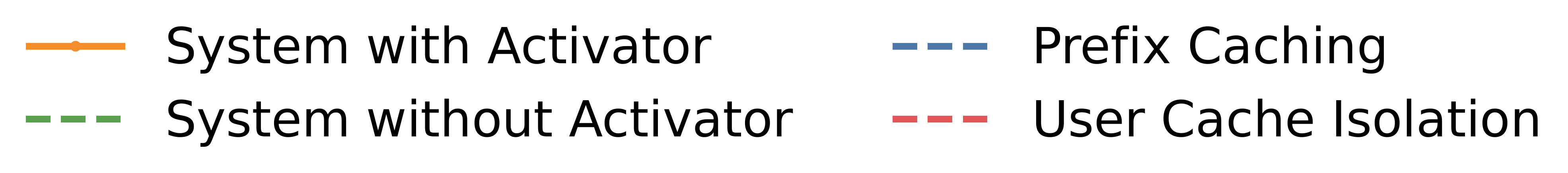}
\end{subfigure}

\begin{subfigure}[t]{\columnwidth}
    \centering
    \includegraphics[width=\linewidth]{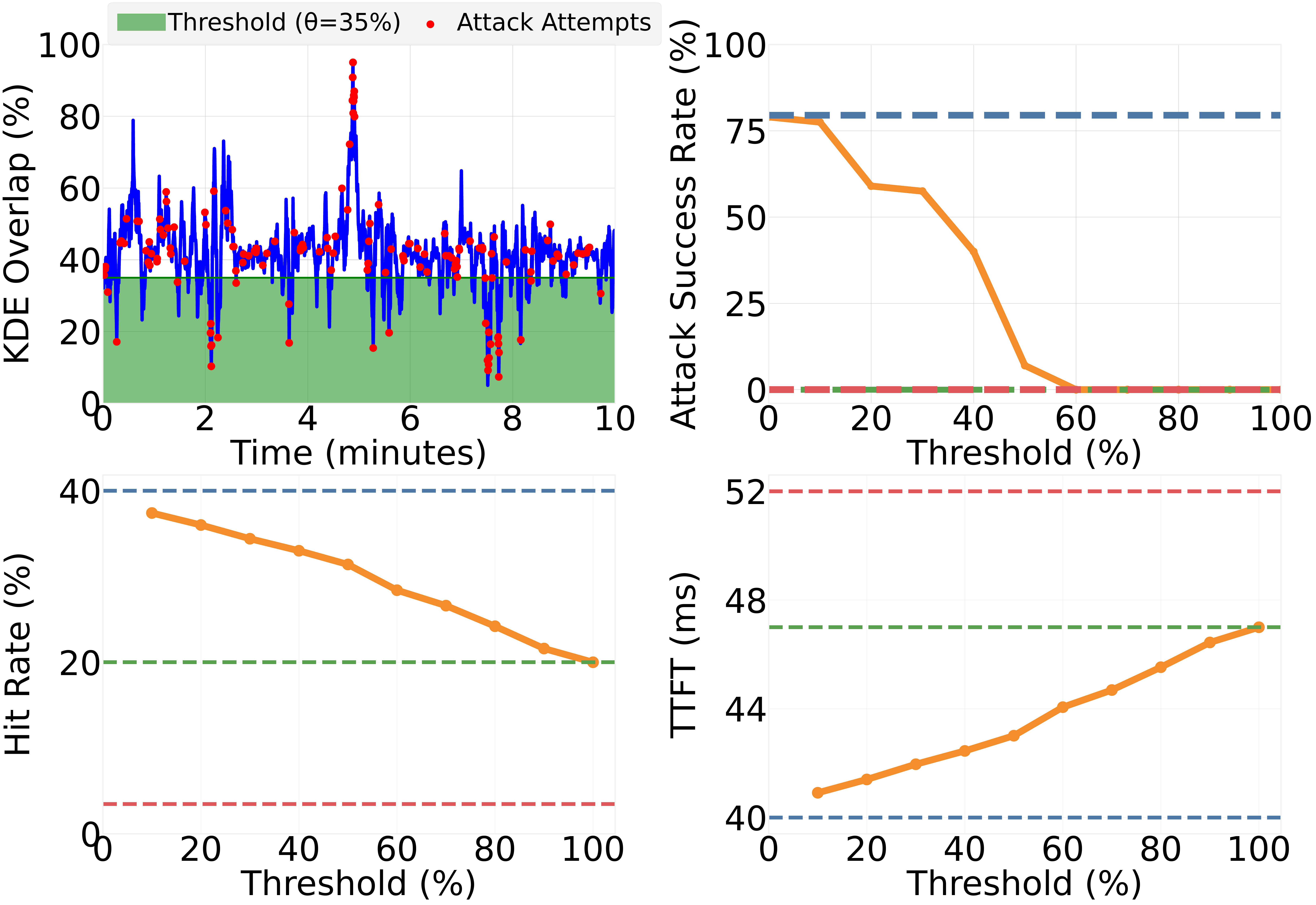}
    % \caption{TTFT vs. hit rate.}
    \label{fig:sensitivity_perf}
\end{subfigure}\hfill

\vspace{-20pt}
\caption{{Sensitivity analysis of the Activator's threshold $\theta$ selection on system security and performance.}}
\label{fig:sensitivity}
\vspace{-0.1in}
\end{figure}

We evaluate \sys{}’s sensitivity to the Activator's threshold $\theta$. Given the KDE overlap over a window $h$, the Activator enables prefix isolation (the Detector) when  overlap($h$) $< \theta$. Thus, $\theta = 0\%$ corresponds to the Prefix Caching baseline (no security) and $\theta = 100\%$ to \sys with the Detector always on. We run Llama2-13B on a single A100 GPU over {\it Workload 4} for 10 minutes. {Figure~\ref{fig:sensitivity} shows how the overlap varies over time, due to the workload's prompt lengths and request rate. We consider a continuous attack setting, marking in red the time points where the attacker successfully reconstructs sensitive information and attempts to steal the prompt.} 

{Figure~\ref{fig:sensitivity} reports the corresponding attack success rate (ASR), capturing how often attacks succeed as a function of both the threshold $\theta$ and the strength of the timing side channel. When $\theta = 0\%$, the ASR is below 100\% because some attacks inherently fail due to weak timing differences (high KDE overlap) regardless of the threshold. As $\theta$ increases, the ASR decreases, since the Activator isolates a larger fraction of prefixes. Intuitively, for a given $\theta$, \sys protects against attacks that fall within the green region in Figure~\ref{fig:sensitivity}. Beyond $\theta \approx 60\%$, \sys reduces the ASR to zero, fully preventing successful prompt stealing attacks.} 
Figure~\ref{fig:sensitivity} shows the corresponding performance (hit rate and average TTFT). As expected, higher $\theta$ leads to more prefix isolation, thus reduced hit rate and increased TTFT, capturing the performance-security trade-off that \sys's Activator introduces.

\compactpara{{Recommendations on the use of prefix isolation and its activation threshold.}} 
{Table~\ref{tab:theta_guidelines} summarizes threshold recommendations for system administrators. For privacy-critical applications (e.g., medical, legal, confidential chats), high thresholds (80--100\%) provide strong protection. For general-purpose applications (e.g., customer support, productivity assistants), mid-range thresholds (50--70\%) balance security and performance. For latency-sensitive or high-throughput systems (e.g., real-time assistants, streaming agents, or high-traffic APIs), lower thresholds (20--40\%) prioritize performance, as high load already limits attack success.}

{Lastly, for privacy-critical applications, even high thresholds may fail to secure a small fraction of prompts, as discussed in \S\ref{sec:security_isolation} and \S\ref{sec:eval:security}. If absolute security is needed, full cache isolation is the only reliable approach. Neither \sys nor SafeKV provide complete security: both leave approximately 3-4\% of prompts unprotected (Figure~\ref{fig:security_datasets}), due to \sys{}’s first-entry, first-try limitation and semantic misclassification in SafeKV, respectively. 
%In such settings, full cache isolation should be preferred, accepting the performance cost when security is the primary objective.
}

\begin{table}[t]
\centering
\small
\setlength{\tabcolsep}{3pt}
\begin{tabular}{l c p{4.5cm}}
\hline
\textbf{Application Type} & \textbf{$\theta$} & \textbf{Rationale} \\
\hline
Privacy-critical & 80-100\% & High security, acceptable performance overheads \\

General-purpose & 50-70\% & Balanced between security and cache reuse, performance \\

Latency-sensitive & 20-40\% & High performance, less isolation as high load already limits attack success \\
\hline
\end{tabular}
\caption{{Recommended threshold $\theta$ by application type.}}
\label{tab:theta_guidelines}
\vspace{-0.3in}
\end{table}
\section{Related Work}
\label{sec:related}

\compactpara{\bf Optimizations of LLM Serving Systems.}
Recent system-level optimizations use {\it approximate} techniques, such as modifying the attention mechanism~\cite{liu2024ringattention, attention-sinks, jiang2024minference, desai2025hashattention, sample-attention}, reusing~\cite{gim2024promptcache, attention-store, cache-blend}, or compressing the KV cache~\cite{infini-gen, fu2024icmlw-lazyllm, sun2025shadowkv, cache-gen, quest, shi-etal-2026-discovering}.
However, these techniques sacrifice accuracy for faster inference. 
In contrast, {\it exact} techniques aim to reduce the latency by efficiently managing the KV cache~\cite{kwon2023pagedattention, qin2025mooncake, pensieve}, sharing prefixes~\cite{chunk-attention, rag-cache, juravsky2024hydragen, sllm}, or optimizing scheduling decisions~\cite{sarathi, sarathi-serve, mnemosyne, deepspeed-fastgen, prepacking, attention-offloading}.
Other approaches attempt to mitigate the impact of long-context processing by disaggregating the prefill and decode phases of inference~\cite{dist-serve, splitwise} or rescheduling requests to reduce memory fragmentation~\cite{llumnix}.

% \thaleia{timing side channels in llms}

\compactpara{\bf Side-channels and performance optimizations.} 
Performance optimizations improve average-case resource usage (e.g., latency, memory, energy), but introduce variability that can expose side channels exploitable by attackers. Historically, attacks targeted physical effects such as power or electromagnetic leakage, requiring direct access. More recently, they exploit shared hardware resources such as caches~\cite{flushreload}, DRAM~\cite{drama}, BPUs~\cite{branchscope}, 
%TLBs~\cite{tlb},
and execution ports~\cite{portcontention}. Side channels also arise in software systems like browsers~\cite{browser1,browser2}, operating systems~\cite{os1}, and network protocols~\cite{network1,network2}.

\compactpara{\bf Timing Side-Channel Attacks in LLM Serving Systems.}
Recent work shows that optimizations such as KV caching and prefix reuse introduce exploitable timing leaks, where differences between cache hits and misses can expose private information in prompts~\cite{early-bird,input-snatch,pmlr-v267-gu25b}. Beyond caching, inference optimizations (e.g., speculative decoding) also introduce timing variations that can leak sensitive information~\cite{remote-attacks}. Other work exploits scheduling behavior rather than timing differences to reconstruct prompts~\cite{prompt-peek}. Together, these works highlight the vulnerabilities introduced by LLM serving optimizations and the need for robust defenses.

\section{Summary}
\label{sec:summary}

We present \sys, a system that secures LLM serving against timing side-channel attacks introduced by Automatic Prefix Caching (APC) without sacrificing performance. \sys monitors cross-user prefix reuse, flags suspicious reuse of cached prefixes, and selectively isolates affected prefixes using lightweight KV cache metadata, without the need to analyze prompt semantics. Our evaluation shows that \sys provides effective protection against prompt-stealing attacks while maintaining high performance and scalability, bridging the gap between insecure APC and slow isolation-based defenses.

\begin{acks}
% \compactpara{Acknowledgements}
We thank the anonymous reviewers for their constructive feedback.
The project has been partially funded by the \grantsponsor{1}{European Research Council}{https://erc.europa.eu/} under the ERC Consolidator grant PRIMULA \grantnum{1}{101230068}.\footnote{Views and opinions expressed are however those of the author(s) only and do not necessarily reflect those of the European Union or the ERC. Neither the European Union nor the ERC can be held responsible for them.}
Also from the Spanish Ministry of Science, Innovation and Universities (MICIU/AEI/10.13039/501100011033) under the projects EUR2025-164828 (SINTRAZAS), RYC2021-032614-I, PID2022-142290 OB-I00 (ESPADA), and CEX2024-001471-M; the Madrid Regional Government through the César Nombela grant (2024-T1/COM-31302) and from the Comunidad de Madrid through the DATIA project, co-funded by FEDER UE.
\end{acks}

% Acknowledgements
%\begin{acks}
%We thank the reviewers for their constructive feedback.\\ 
%This work is part of the grant FJC2021-047102-I, funded by MCIN/AEI/10.13039/501100011033 and the European Union «NextGenerationEU»/PRTR.
%\end{acks}

% Bibliography style and file
\bibliographystyle{ACM-Reference-Format}
\bibliography{paper}

@misc{sglang,
  title={SGLang: Efficient Serving for Large Language Models},
  author={SGLang Team},
  howpublished={\url{https://github.com/sgl-project/sglang}},
  year={2024}
}

@misc{vllm_prefix_caching,
  title={Automatic Prefix Caching in vLLM},
  author={vLLM Team},
  howpublished={\url{https://docs.vllm.ai/en/latest/features/automatic_prefix_caching/}},
  year={2025}
}

@article{brown2020language,
  title={Language Models are Few-Shot Learners},
  author={Brown, Tom and Mann, Benjamin and Ryder, Nick and others},
  journal={Advances in Neural Information Processing Systems},
  volume={33},
  pages={1877--1901},
  year={2020}
}

@article{touvron2023llama,
  title={LLaMA: Open and Efficient Foundation Language Models},
  author={Touvron, Hugo and others},
  journal={arXiv preprint arXiv:2302.13971},
  year={2023}
}

@misc{vllm,
  title={vLLM: High-Throughput Serving for Large Language Models},
  author={vLLM Team},
  howpublished={\url{https://github.com/vllm-project/vllm}},
  year={2024}
}

@inproceedings{gras_aslr_2017,
  title     = {{ASLR} on the Line: Practical Cache Attacks on the {MMU}},
  author    = {Gras, Ben and Razavi, Kaveh and Bosman, Erik and Bos, Herbert and Giuffrida, Cristiano},
  booktitle = {Proceedings of the Network and Distributed System Security Symposium (NDSS '17)},
  year      = {2017},
  url       = {https://download.vusec.net/papers/anc_ndss17.pdf}
}

@inproceedings {browser1,
author = {Pepe Vila and Boris Kopf},
title = {Loophole: Timing Attacks on Shared Event Loops in Chrome},
booktitle = {26th USENIX Security Symposium (USENIX Security 17)},
year = {2017},
isbn = {978-1-931971-40-9},
address = {Vancouver, BC},
pages = {849--864},
url = {https://www.usenix.org/conference/usenixsecurity17/technical-sessions/presentation/vila},
publisher = {USENIX Association},
month = aug
}

@inproceedings{browser2,
author = {Snyder, Peter and Karami, Soroush and Edelstein, Arthur and Livshits, Benjamin and Haddadi, Hamed},
title = {Pool-party: exploiting browser resource pools for web tracking},
year = {2023},
isbn = {978-1-939133-37-3},
publisher = {USENIX Association},
address = {USA},
booktitle = {Proceedings of the 32nd USENIX Conference on Security Symposium},
articleno = {397},
numpages = {15},
location = {Anaheim, CA, USA},
series = {SEC '23}
}

@inproceedings {network1,
author = {Aastha Mehta and Mohamed Alzayat and Roberta De Viti and Bj{\"o}rn B. Brandenburg and Peter Druschel and Deepak Garg},
title = {Pacer: Comprehensive Network {Side-Channel} Mitigation in the Cloud},
booktitle = {31st USENIX Security Symposium (USENIX Security 22)},
year = {2022},
isbn = {978-1-939133-31-1},
address = {Boston, MA},
pages = {2819--2838},
url = {https://www.usenix.org/conference/usenixsecurity22/presentation/mehta},
publisher = {USENIX Association},
month = aug
}

@INPROCEEDINGS{network2,

  author={Wright, Charles V. and Ballard, Lucas and Coull, Scott E. and Monrose, Fabian and Masson, Gerald M.},

  booktitle={2008 IEEE Symposium on Security and Privacy (sp 2008)}, 

  title={Spot Me if You Can: Uncovering Spoken Phrases in Encrypted VoIP Conversations}, 

  year={2008},

  volume={},

  number={},

  pages={35-49},

  doi={10.1109/SP.2008.21}}

@inproceedings{os1,
author = {H\"{a}hnel, Marcus and Cui, Weidong and Peinado, Marcus},
title = {High-resolution side channels for untrusted operating systems},
year = {2017},
isbn = {9781931971386},
publisher = {USENIX Association},
address = {USA},
booktitle = {Proceedings of the 2017 USENIX Conference on Usenix Annual Technical Conference},
pages = {299–312},
numpages = {14},
location = {Santa Clara, CA, USA},
series = {USENIX ATC '17}
}

@inproceedings{flushreload,
author={Yuval Yarom and Katrina Falkner},
title={{FLUSH+RELOAD}: A High Resolution, Low Noise, L3 Cache Side-Channel Attack},
booktitle={Proceedings of the 23rd {USENIX} Security Symposium}, 
series = { {USENIX} Security '14},
publisher={{USENIX} Association},
year={2014}
}

@inproceedings {dnscache,
author = {Rasmus Dahlberg and Tobias Pulls},
title = {Timeless Timing Attacks and Preload Defenses in Tor{\textquoteright}s {DNS} Cache},
booktitle = {32nd USENIX Security Symposium (USENIX Security 23)},
year = {2023},
isbn = {978-1-939133-37-3},
address = {Anaheim, CA},
pages = {2635--2652},
url = {https://www.usenix.org/conference/usenixsecurity23/presentation/dahlberg},
publisher = {USENIX Association},
month = aug
}

@inproceedings{10.5555/3241094.3241131, author = {Kohlbrenner, David and Shacham, Hovav}, title = {Trusted browsers for uncertain times}, year = {2016}, isbn = {9781931971324}, publisher = {USENIX Association}, address = {USA}, booktitle = {Proceedings of the 25th USENIX Conference on Security Symposium}, pages = {463–480}, numpages = {18}, location = {Austin, TX, USA}, series = {SEC'16} }

@inproceedings{browsercache, author = {Bortz, Andrew and Boneh, Dan}, title = {Exposing private information by timing web applications}, year = {2007}, isbn = {9781595936547}, publisher = {Association for Computing Machinery}, address = {New York, NY, USA}, url = {https://doi.org/10.1145/1242572.1242656}, doi = {10.1145/1242572.1242656}, booktitle = {Proceedings of the 16th International Conference on World Wide Web}, pages = {621–628}, numpages = {8}, location = {Banff, Alberta, Canada}, series = {WWW '07} }

@inproceedings{pagecache, author = {Gruss, Daniel and Kraft, Erik and Tiwari, Trishita and Schwarz, Michael and Trachtenberg, Ari and Hennessey, Jason and Ionescu, Alex and Fogh, Anders}, title = {Page Cache Attacks}, year = {2019}, isbn = {9781450367479}, publisher = {Association for Computing Machinery}, address = {New York, NY, USA}, url = {https://doi.org/10.1145/3319535.3339809}, doi = {10.1145/3319535.3339809}, booktitle = {Proceedings of the 2019 ACM SIGSAC Conference on Computer and Communications Security}, pages = {167–180}, numpages = {14}, location = {London, United Kingdom}, series = {CCS '19} }

@inproceedings{drama,
author={Peter Pessl and Daniel Gruss and Cl\'ementine Maurice and Michael Schwarz and Stefan Mangard},
title={{DRAMA}: Exploiting {DRAM} Addressing for Cross-{CPU} Attacks},
booktitle={Proceedings of the 25th {USENIX} Security Symposium}, 
series = { {USENIX} Security '16},
publisher={{USENIX} Association},
year={2016}
}

@inproceedings{branchscope,
author={Dmitry Evtyushkin and Ryan Riley and Nael B. Abu-Ghazaleh and Dmitry Ponomarev},
title={{BranchScope: A New Side-Channel Attack on Directional Branch Predictor}},
booktitle={Proceedings of the 23rd International Conference on Architectural Support for Programming Languages and Operating Systems}, 
series={{ASPLOS} '18},
publisher={ACM},
year={2018}
}

@inproceedings{portcontention,
author={Alejandro Cabrera Aldaya and Billy Bob Brumley and Sohaib ul Hassan and Cesar Pereida Garc\'ia and Nicola Tuveri},
title={{Port Contention for Fun and Profit}},
booktitle={Proceedings of the 40th IEEE Symposium on Security and Privacy}, 
publisher={IEEE},
series={{S\&P} '19},
year={2019}
}

@misc{chu2026selectivekvcachesharingmitigate,
      title={Selective KV-Cache Sharing to Mitigate Timing Side-Channels in LLM Inference}, 
      author={Kexin Chu and Zecheng Lin and Dawei Xiang and Zixu Shen and Jianchang Su and Cheng Chu and Yiwei Yang and Wenhui Zhang and Wenfei Wu and Wei Zhang},
      year={2026},
      eprint={2508.08438},
      archivePrefix={arXiv},
      primaryClass={cs.CR},
      url={https://arxiv.org/abs/2508.08438}, 
}

@misc{google_dp,
  author       = {{Google Research}},
  title        = {Securing Private Data at Scale with Differentially Private Partition Selection},
  howpublished = {\url{https://research.google/blog/securing-private-data-at-scale-with-differentially-private-partition-selection/}},
  year         = {2022},
  note         = {Accessed: 2026-05}
}

@ARTICLE{early-bird,
  author={Song, Linke and Pang, Zixuan and Wang, Wenhao and Wang, Zihao and Wang, XiaoFeng and Chen, Hongbo and Song, Wei and Jin, Yier and Meng, Dan and Hou, Rui},
  journal={IEEE Transactions on Information Forensics and Security}, 
  title={The Early Bird Catches the Leak: Unveiling Timing Side Channels in LLM Serving Systems}, 
  year={2025},
  volume={20},
  number={},
  pages={11431-11446},
  doi={10.1109/TIFS.2025.3622954}
  }

@inproceedings{prompt-peek,
  title={I Know What You Asked: Prompt Leakage via KV-Cache Sharing in Multi-Tenant LLM Serving},
  author={Wu, Guanlong and Zhang, Zheng and Zhang, Yao and Wang, Weili and Niu, Jianyu and Wu, Ye and Zhang, Yinqian},
  booktitle={Proceedings of the 2025 Network and Distributed System Security (NDSS) Symposium. San Diego, CA, USA},
  year={2025}
}

@INPROCEEDINGS{cache-partitioning,
  author={Pang, Zixuan and Wang, Wenhao and Liao, Yong},
  booktitle={2024 6th International Conference on Frontier Technologies of Information and Computer (ICFTIC)}, 
  title={Cache Partitioning for Mitigating Timing Side-Channel Attacks in LLM Serving Systems}, 
  year={2024},
  volume={},
  number={},
  pages={1238-1245},
  doi={10.1109/ICFTIC64248.2024.10913329}}

@misc{input-snatch,
      title={InputSnatch: Stealing Input in LLM Services via Timing Side-Channel Attacks}, 
      author={Xinyao Zheng and Husheng Han and Shangyi Shi and Qiyan Fang and Zidong Du and Xing Hu and Qi Guo},
      year={2024},
      eprint={2411.18191},
      archivePrefix={arXiv},
      primaryClass={cs.CR},
      url={https://arxiv.org/abs/2411.18191}, 
}

@misc{remote-attacks,
      title={Remote Timing Attacks on Efficient Language Model Inference}, 
      author={Nicholas Carlini and Milad Nasr},
      year={2024},
      eprint={2410.17175},
      archivePrefix={arXiv},
      primaryClass={cs.CR},
      url={https://arxiv.org/abs/2410.17175}, 
}

@inproceedings{konpap-euromlsys24,
author = {Papaioannou, Konstantinos and Doudali, Thaleia Dimitra},
title = {The Importance of Workload Choice in Evaluating LLM Inference Systems},
year = {2024},
isbn = {9798400705410},
publisher = {Association for Computing Machinery},
address = {New York, NY, USA},
url = {https://doi.org/10.1145/3642970.3655823},
doi = {10.1145/3642970.3655823},
booktitle = {Proceedings of the 4th Workshop on Machine Learning and Systems},
pages = {39–46},
numpages = {8},
location = {Athens, Greece},
series = {EuroMLSys '24}
}

@online{gemma3_4b_it,
  author       = {{Google DeepMind}},
  title        = {Gemma 3 4B Instruct Model},
  year         = {2025},
  url          = {https://huggingface.co/google/gemma-3-4b-it},
  note         = {Available on Hugging Face Hub}
}

@online{llava_onevision_qwen2_0_5b_ov_hf,
  author       = {{LLaVA Team}},
  title        = {LLaVA-OneVision Qwen2 0.5B (OV-HF)},
  year         = {2024},
  url          = {https://huggingface.co/llava-hf/llava-onevision-qwen2-0.5b-ov-hf},
  note         = {Available on Hugging Face Hub}
}

@online{llava_onevision_qwen2_7b_ov_chat_hf,
  author       = {{LLaVA Team}},
  title        = {LLaVA-OneVision Qwen2 7B OV-Chat (HF)},
  year         = {2024},
  url          = {https://huggingface.co/llava-hf/llava-onevision-qwen2-7b-ov-chat-hf},
  note         = {Available on Hugging Face Hub}
}

@online{llama2_7b_chat_hf,
  author       = {{Meta AI}},
  title        = {Llama 2 7B Chat Model},
  year         = {2023},
  url          = {https://huggingface.co/meta-llama/Llama-2-7b-chat-hf},
  note         = {Available on Hugging Face Hub}
}

@online{llama2_13b_chat_hf,
  author       = {{Meta AI}},
  title        = {Llama 2 13B Chat Model},
  year         = {2023},
  url          = {https://huggingface.co/meta-llama/Llama-2-13b-chat-hf},
  note         = {Available on Hugging Face Hub}
}

@online{qwen2_vl_2b_instruct,
  author       = {{Alibaba Cloud}},
  title        = {Qwen2-VL 2B Instruct Model},
  year         = {2024},
  url          = {https://huggingface.co/Qwen/Qwen2-VL-2B-Instruct},
  note         = {Available on Hugging Face Hub}
}

@online{qwen2_5_vl_3b_instruct,
  author       = {{Alibaba Cloud}},
  title        = {Qwen2.5-VL 3B Instruct Model},
  year         = {2024},
  url          = {https://huggingface.co/Qwen/Qwen2.5-VL-3B-Instruct},
  note         = {Available on Hugging Face Hub}
}

@online{qwen2_vl_7b_instruct,
  author       = {{Alibaba Cloud}},
  title        = {Qwen2-VL 7B Instruct Model},
  year         = {2024},
  url          = {https://huggingface.co/Qwen/Qwen2-VL-7B-Instruct},
  note         = {Available on Hugging Face Hub}
}

@online{qwen2_5_vl_7b_instruct,
  author       = {{Alibaba Cloud}},
  title        = {Qwen2.5-VL 7B Instruct Model},
  year         = {2024},
  url          = {https://huggingface.co/Qwen/Qwen2.5-VL-7B-Instruct},
  note         = {Available on Hugging Face Hub}
}

@online{deepmind2025gemini,
  author       = {DeepMind},
  title        = {Gemini},
  year         = {2025},
  url          = {https://deepmind.google/technologies/gemini/}
}

@online{deepseek2024contextcaching,
  author       = {DeepSeek},
  title        = {DeepSeek API Docs: DeepSeek API Introduces Context Caching on Disk, Cutting Prices by an Order of Magnitude},
  year         = {2024},
  note         = {Accessed: 2025-07-17},
  url          = {https://api-docs.deepseek.com/news/news0802/}
}

@inproceedings{kwon2023pagedattention,
  author    = {Kwon, Woosuk and Li, Zhuohan and Zhuang, Siyuan and Sheng, Ying and Zheng, Lianmin and Yu, Cody Hao and Gonzalez, Joseph E. and Zhang, Hao and Stoica, Ion},
  title     = {Efficient Memory Management for Large Language Model Serving with {PagedAttention}},
  booktitle = {Proceedings of the 29th Symposium on Operating Systems Principles (SOSP '23)},
  year      = {2023},
  pages     = {611--626},
  doi       = {10.1145/3600006.3613165}
}

@online{openai2024promptcaching,
  author       = {OpenAI},
  title        = {Prompt Caching: Reduce Latency and Cost with Prompt Caching},
  year         = {2024},
  note         = {Accessed: 2025-07-17},
  url          = {https://platform.openai.com/docs/guides/prompt-caching}
}

@inproceedings{qin2025mooncake,
  author    = {Qin, Ruoyu and Li, Zheming and He, Weiran and Cui, Jialei and Ren, Feng and Zhang, Mingxing and Wu, Yongwei and Zheng, Weimin and Xu, Xinran},
  title     = {Mooncake: Trading More Storage for Less Computation --- A {KVCache}-Centric Architecture for Serving {LLM} Chatbot},
  booktitle = {23rd USENIX Conference on File and Storage Technologies (FAST '25)},
  year      = {2025},
  pages     = {155--170}
}

@inproceedings{zheng2024sglang,
  author    = {Zheng, Lianmin and Yin, Liangsheng and Xie, Zhiqiang and Sun, Chuyue and Huang, Jeff and Yu, Cody Hao and Cao, Shiyi and Kozyrakis, Christos and Stoica, Ion and Gonzalez, Joseph E. and Barrett, Clark and Sheng, Ying},
  title     = {{SGLang}: Efficient Execution of Structured Language Model Programs},
  booktitle = {Advances in Neural Information Processing Systems (NeurIPS '24)},
  volume    = {37},
  pages     = {62557--62583},
  year      = {2024}
}

@misc{anon8231489123_2023,
  author       = {{anon8231489123}},
  title        = {{ShareGPT Vicuna unfiltered}},
  year         = {2023},
  howpublished = {\url{https://huggingface.co/datasets/anon8231489123/ShareGPT\%20Vicuna\%20unfiltered}},
  note         = {Dataset on Hugging Face}
}

@inproceedings{liu2024ringattention,
  title     = {Ring Attention with Blockwise Transformers for Near-Infinite Context},
  author    = {Liu, Hao and Zaharia, Matei and Abbeel, Pieter},
  booktitle = {International Conference on Learning Representations (ICLR)},
  year      = {2024}
}

@misc{attention-sinks,
    title={Efficient Streaming Language Models with Attention Sinks}, 
    author={Guangxuan Xiao and Yuandong Tian and Beidi Chen and Song Han and Mike Lewis},
    year={2024},
    eprint={2309.17453},
    archivePrefix={arXiv},
    primaryClass={cs.CL},
    url={https://arxiv.org/abs/2309.17453}, 
}

@inproceedings{jiang2024minference,
  title     = {MInference 1.0: Accelerating Pre-filling for Long-Context LLMs via Dynamic Sparse Attention},
  author    = {Jiang, Huiqiang and Li, Yucheng and Zhang, Chengruidong and Wu, Qianhui and Luo, Xufang and Ahn, Surin and Han, Zhenhua and Abdi, Amir H. and Li, Dongsheng and Lin, Chin-Yew and Yang, Yuqing and Qiu, Lili},
  booktitle = {Advances in Neural Information Processing Systems (NeurIPS)},
  volume    = {37},
  year      = {2024}
}

@inproceedings{desai2025hashattention,
  title     = {HashAttention: Semantic Sparsity for Faster Inference},
  author    = {Desai, Aditya and Yang, Shuo and Cuadron, Alejandro and Klimovic, Ana and Zaharia, Matei and Gonzalez, Joseph E. and Stoica, Ion},
  booktitle = {Proceedings of the 42nd International Conference on Machine Learning (ICML)},
  series    = {PMLR},
  volume    = {267},
  year      = {2025}
}

@inproceedings{sample-attention,
    title={SampleAttention: Near-Lossless Acceleration of Long Context {LLM} Inference with Adaptive Structured Sparse Attention},
    author={Qianchao Zhu and Jiangfei Duan and Chang Chen and Siran Liu and Xiuhong Li and Guanyu Feng and Xin Lv and Xiao Chuanfu and Dahua Lin and Chao Yang},
    booktitle={Eighth Conference on Machine Learning and Systems},
    year={2025},
    url={https://openreview.net/forum?id=RuZ80yl71h}
}

@inproceedings{attention-store,
    author = {Gao, Bin and He, Zhuomin and Sharma, Puru and Kang, Qingxuan and Jevdjic, Djordje and Deng, Junbo and Yang, Xingkun and Yu, Zhou and Zuo, Pengfei},
    title = {Cost-efficient large language model serving for multi-turn conversations with CachedAttention},
    year = {2024},
    isbn = {978-1-939133-41-0},
    publisher = {USENIX Association},
    address = {USA},
    booktitle = {Proceedings of the 2024 USENIX Conference on Usenix Annual Technical Conference},
    articleno = {7},
    numpages = {16},
    location = {Santa Clara, CA, USA},
    series = {USENIX ATC'24}
}

@inproceedings{cache-blend,
    author = {Yao, Jiayi and Li, Hanchen and Liu, Yuhan and Ray, Siddhant and Cheng, Yihua and Zhang, Qizheng and Du, Kuntai and Lu, Shan and Jiang, Junchen},
    title = {CacheBlend: Fast Large Language Model Serving for RAG with Cached Knowledge Fusion},
    year = {2025},
    isbn = {9798400711961},
    publisher = {Association for Computing Machinery},
    address = {New York, NY, USA},
    url = {https://doi.org/10.1145/3689031.3696098},
    doi = {10.1145/3689031.3696098},
    booktitle = {Proceedings of the Twentieth European Conference on Computer Systems},
    pages = {94–109},
    numpages = {16},
    location = {Rotterdam, Netherlands},
    series = {EuroSys '25}
}

@inproceedings{gim2024promptcache,
  title     = {Prompt Cache: Modular Attention Reuse for Low-Latency Inference},
  author    = {Gim, In and Chen, Guojun and Lee, Seung-seob and Sarda, Nikhil and Khandelwal, Anurag and Zhong, Lin},
  booktitle = {Proceedings of Machine Learning and Systems (MLSys)},
  year      = {2024}
}

@inproceedings {infini-gen,
    author = {Wonbeom Lee and Jungi Lee and Junghwan Seo and Jaewoong Sim},
    title = {{InfiniGen}: Efficient Generative Inference of Large Language Models with Dynamic {KV} Cache Management},
    booktitle = {18th USENIX Symposium on Operating Systems Design and Implementation (OSDI 24)},
    year = {2024},
    isbn = {978-1-939133-40-3},
    address = {Santa Clara, CA},
    pages = {155--172},
    url = {https://www.usenix.org/conference/osdi24/presentation/lee},
    publisher = {USENIX Association},
    month = jul
}

@inproceedings{fu2024icmlw-lazyllm,
  title     = {{LazyLLM: Dynamic Token Pruning for Efficient Long Context LLM Inference}},
  author    = {Fu, Qichen and Cho, Minsik and Merth, Thomas and Mehta, Sachin and Rastegari, Mohammad and Najibi, Mahyar},
  booktitle = {ICML 2024 Workshops: ES-FoMo-II},
  year      = {2024},
  url       = {https://mlanthology.org/icmlw/2024/fu2024icmlw-lazyllm/}
}

@inproceedings{cache-gen,
    author = {Liu, Yuhan and Li, Hanchen and Cheng, Yihua and Ray, Siddhant and Huang, Yuyang and Zhang, Qizheng and Du, Kuntai and Yao, Jiayi and Lu, Shan and Ananthanarayanan, Ganesh and Maire, Michael and Hoffmann, Henry and Holtzman, Ari and Jiang, Junchen},
    title = {CacheGen: KV Cache Compression and Streaming for Fast Large Language Model Serving},
    year = {2024},
    isbn = {9798400706141},
    publisher = {Association for Computing Machinery},
    address = {New York, NY, USA},
    url = {https://doi.org/10.1145/3651890.3672274},
    doi = {10.1145/3651890.3672274},
    booktitle = {Proceedings of the ACM SIGCOMM 2024 Conference},
    pages = {38–56},
    numpages = {19},
    location = {Sydney, NSW, Australia},
    series = {ACM SIGCOMM '24}
}

@inproceedings{sun2025shadowkv,
  title     = {ShadowKV: KV Cache in Shadows for High-Throughput Long-Context LLM Inference},
  author    = {Sun, Hanshi and Chang, Li-Wen and Bao, Wenlei and Zheng, Size and Zheng, Ningxin and Liu, Xin and Dong, Harry and Chi, Yuejie and Chen, Beidi},
  booktitle = {Proceedings of the 42nd International Conference on Machine Learning (ICML)},
  series    = {PMLR},
  volume    = {267},
  year      = {2025}
}

@inproceedings{quest,
    author = {Tang, Jiaming and Zhao, Yilong and Zhu, Kan and Xiao, Guangxuan and Kasikci, Baris and Han, Song},
    title = {QUEST: query-aware sparsity for efficient long-context LLM inference},
    year = {2024},
    publisher = {JMLR.org},
    booktitle = {Proceedings of the 41st International Conference on Machine Learning},
    articleno = {1955},
    numpages = {11},
    location = {Vienna, Austria},
    series = {ICML'24}
}

@inproceedings{shi-etal-2026-discovering,
    title     = {Discovering the Gems in Early Layers: Accelerating Long-Context {LLM}s with 1000x Input Token Reduction},
    author    = {Shi, Zhenmei and Ming, Yifei and Nguyen, Xuan-Phi and Liang, Yingyu and Joty, Shafiq},
    booktitle = {Findings of the Association for Computational Linguistics: ACL 2026},
    year      = {2026},
    pages     = {13839--13857},
    address   = {San Diego, California, United States},
    publisher = {Association for Computational Linguistics},
    doi       = {10.18653/v1/2026.findings-acl.677}
}

@inproceedings{pensieve,
    author = {Yu, Lingfan and Lin, Jinkun and Li, Jinyang},
    title = {Stateful Large Language Model Serving with Pensieve},
    year = {2025},
    isbn = {9798400711961},
    publisher = {Association for Computing Machinery},
    address = {New York, NY, USA},
    url = {https://doi.org/10.1145/3689031.3696086},
    doi = {10.1145/3689031.3696086},
    booktitle = {Proceedings of the Twentieth European Conference on Computer Systems},
    pages = {144–158},
    numpages = {15},
    location = {Rotterdam, Netherlands},
    series = {EuroSys '25}
}

@inproceedings{chunk-attention,
    title = "{C}hunk{A}ttention: Efficient Self-Attention with Prefix-Aware {KV} Cache and Two-Phase Partition",
    author = "Ye, Lu  and
      Tao, Ze  and
      Huang, Yong  and
      Li, Yang",
    editor = "Ku, Lun-Wei  and
      Martins, Andre  and
      Srikumar, Vivek",
    booktitle = "Proceedings of the 62nd Annual Meeting of the Association for Computational Linguistics (Volume 1: Long Papers)",
    month = aug,
    year = "2024",
    address = "Bangkok, Thailand",
    publisher = "Association for Computational Linguistics",
    url = "https://aclanthology.org/2024.acl-long.623/",
    doi = "10.18653/v1/2024.acl-long.623",
    pages = "11608--11620"
}

@inproceedings{sllm,
    author={Lin, Jieyu and Zhang, Sai Qian and Leon-Garcia, Alberto},
    booktitle={2024 25th International Symposium on Quality Electronic Design (ISQED)}, 
    title={sLLM: Accelerating LLM Inference using Semantic Load Balancing with Shared Memory Data Structures}, 
    year={2024},
    volume={},
    number={},
    pages={1-6},
    doi={10.1109/ISQED60706.2024.10528703}
}

@article{rag-cache,
    author = {Jin, Chao and Zhang, Zili and Jiang, Xuanlin and Liu, Fangyue and Liu, Shufan and Liu, Xuanzhe and Jin, Xin},
    title = {RAGCache: Efficient Knowledge Caching for Retrieval-Augmented Generation},
    year = {2025},
    issue_date = {February 2026},
    publisher = {Association for Computing Machinery},
    address = {New York, NY, USA},
    volume = {44},
    number = {1},
    issn = {0734-2071},
    url = {https://doi.org/10.1145/3768628},
    doi = {10.1145/3768628},
    journal = {ACM Trans. Comput. Syst.},
    month = nov,
    articleno = {2},
    numpages = {27}
}

@article{kde-overlap,
  title={The overlapping coefficient as a measure of agreement between probability distributions},
  author={Inman, Henry F. and Bradley, Edwin L.},
  journal={Communications in Statistics-Theory and Methods},
  volume={18},
  number={10},
  pages={3851--3874},
  year={1989},
  publisher={Taylor \& Francis},
  doi={10.1080/03610928908830127}
}

@inproceedings{
juravsky2024hydragen,
title={Hydragen: High-Throughput {LLM} Inference with Shared Prefixes},
author={Jordan Juravsky and Bradley Brown and Ryan Saul Ehrlich and Daniel Y Fu and Christopher Re and Azalia Mirhoseini},
booktitle={Workshop on Efficient Systems for Foundation Models II @ ICML2024},
year={2024},
url={https://openreview.net/forum?id=Ye3ia34Fpp}
}

@misc{sarathi,
    title={SARATHI: Efficient LLM Inference by Piggybacking Decodes with Chunked Prefills}, 
    author={Amey Agrawal and Ashish Panwar and Jayashree Mohan and Nipun Kwatra and Bhargav S. Gulavani and Ramachandran Ramjee},
    year={2023},
    eprint={2308.16369},
    archivePrefix={arXiv},
    primaryClass={cs.LG},
    url={https://arxiv.org/abs/2308.16369}, 
}

@inproceedings{sarathi-serve,
    author = {Agrawal, Amey and Kedia, Nitin and Panwar, Ashish and Mohan, Jayashree and Kwatra, Nipun and Gulavani, Bhargav S. and Tumanov, Alexey and Ramjee, Ramachandran},
    title = {Taming throughput-latency tradeoff in LLM inference with sarathi-serve},
    year = {2024},
    isbn = {978-1-939133-40-3},
    publisher = {USENIX Association},
    address = {USA},
    booktitle = {Proceedings of the 18th USENIX Conference on Operating Systems Design and Implementation},
    articleno = {7},
    numpages = {18},
    location = {Santa Clara, CA, USA},
    series = {OSDI'24}
}

@misc{mnemosyne,
    title={Mnemosyne: Parallelization Strategies for Efficiently Serving Multi-Million Context Length LLM Inference Requests Without Approximations}, 
    author={Amey Agrawal and Junda Chen and Íñigo Goiri and Ramachandran Ramjee and Chaojie Zhang and Alexey Tumanov and Esha Choukse},
    year={2024},
    eprint={2409.17264},
    archivePrefix={arXiv},
    primaryClass={cs.LG},
    url={https://arxiv.org/abs/2409.17264}, 
}

@misc{prepacking,
    title={Prepacking: A Simple Method for Fast Prefilling and Increased Throughput in Large Language Models}, 
    author={Siyan Zhao and Daniel Israel and Guy Van den Broeck and Aditya Grover},
    year={2024},
    eprint={2404.09529},
    archivePrefix={arXiv},
    primaryClass={cs.LG},
    url={https://arxiv.org/abs/2404.09529}, 
}

@misc{attention-offloading,
    title={Efficient and Economic Large Language Model Inference with Attention Offloading}, 
    author={Shaoyuan Chen and Yutong Lin and Mingxing Zhang and Yongwei Wu},
    year={2024},
    eprint={2405.01814},
    archivePrefix={arXiv},
    primaryClass={cs.LG},
    url={https://arxiv.org/abs/2405.01814}, 
}

@misc{deepspeed-fastgen,
    title={DeepSpeed-FastGen: High-throughput Text Generation for LLMs via MII and DeepSpeed-Inference}, 
    author={Connor Holmes and Masahiro Tanaka and Michael Wyatt and Ammar Ahmad Awan and Jeff Rasley and Samyam Rajbhandari and Reza Yazdani Aminabadi and Heyang Qin and Arash Bakhtiari and Lev Kurilenko and Yuxiong He},
    year={2024},
    eprint={2401.08671},
    archivePrefix={arXiv},
    primaryClass={cs.PF},
    url={https://arxiv.org/abs/2401.08671}, 
}

@inproceedings {dist-serve,
    author = {Yinmin Zhong and Shengyu Liu and Junda Chen and Jianbo Hu and Yibo Zhu and Xuanzhe Liu and Xin Jin and Hao Zhang},
    title = {{DistServe}: Disaggregating Prefill and Decoding for Goodput-optimized Large Language Model Serving},
    booktitle = {18th USENIX Symposium on Operating Systems Design and Implementation (OSDI 24)},
    year = {2024},
    isbn = {978-1-939133-40-3},
    address = {Santa Clara, CA},
    pages = {193--210},
    url = {https://www.usenix.org/conference/osdi24/presentation/zhong-yinmin},
    publisher = {USENIX Association},
    month = jul
}

@inproceedings{splitwise,
    author={Patel, Pratyush and Choukse, Esha and Zhang, Chaojie and Shah, Aashaka and Goiri, Íñigo and Maleki, Saeed and Bianchini, Ricardo},
    booktitle={2024 ACM/IEEE 51st Annual International Symposium on Computer Architecture (ISCA)},
    title={Splitwise: Efficient Generative LLM Inference Using Phase Splitting}, 
    year={2024},
    volume={},
    number={},
    pages={118-132},
    doi={10.1109/ISCA59077.2024.00019}
}

@inproceedings {llumnix,
    author = {Biao Sun and Ziming Huang and Hanyu Zhao and Wencong Xiao and Xinyi Zhang and Yong Li and Wei Lin},
    title = {Llumnix: Dynamic Scheduling for Large Language Model Serving},
    booktitle = {18th USENIX Symposium on Operating Systems Design and Implementation (OSDI 24)},
    year = {2024},
    isbn = {978-1-939133-40-3},
    address = {Santa Clara, CA},
    pages = {173--191},
    url = {https://www.usenix.org/conference/osdi24/presentation/sun-biao},
    publisher = {USENIX Association},
    month = jul
}

@inproceedings{devlin2019bert,
  title={BERT: Pre-training of Deep Bidirectional Transformers for Language Understanding},
  author={Devlin, Jacob and Chang, Ming-Wei and Lee, Kenton and Toutanova, Kristina},
  booktitle={Proceedings of the 2019 Conference of the North American Chapter of the Association for Computational Linguistics: Human Language Technologies, Volume 1 (Long and Short Papers)},
  pages={4171--4186},
  year={2019},
  doi={10.18653/v1/N19-1423}
}

@inproceedings{bang2023gptcache,
  author    = {Bang, Fu},
  title     = {{GPTCache}: An Open-Source Semantic Cache for {LLM} Applications Enabling Faster Answers and Cost Savings},
  booktitle = {Proceedings of the 3rd Workshop for Natural Language Processing Open Source Software (NLP-OSS 2023)},
  year      = {2023},
  pages     = {212--218}
}

@misc{anthropic_caching_docs,
  author       = {{Anthropic}},
  title        = {Prompt Caching --- {Claude} {API} Documentation},
  year         = {2026},
  howpublished = {\url{https://docs.anthropic.com/en/docs/build-with-claude/prompt-caching}},
  note         = {Workspace-level isolation enforced from February 5, 2026}
}

@misc{azure_openai_docs,
  author       = {{Microsoft Azure}},
  title        = {Azure {OpenAI} Service --- Data Privacy and Prompt Caching},
  year         = {2025},
  howpublished = {\url{https://learn.microsoft.com/en-us/azure/ai-services/openai/}},
}

@misc{google_vertex_caching,
  author       = {{Google Cloud}},
  title        = {Context Caching --- {Gemini} {API} Documentation},
  year         = {2025},
  howpublished = {\url{https://cloud.google.com/vertex-ai/generative-ai/docs/context-cache/context-cache-overview}},
}

@misc{amazon_bedrock_docs,
  author       = {{Amazon Web Services}},
  title        = {Prompt Caching for {Amazon Bedrock}},
  year         = {2025},
  howpublished = {\url{https://docs.aws.amazon.com/bedrock/latest/userguide/prompt-caching.html}},
}

@misc{vllm_cache_salt_pr,
  author       = {{vLLM Contributors}},
  title        = {Prevent side-channel attacks via cache salting --- {Pull Request \#17045}},
  year         = {2025},
  howpublished = {\url{https://github.com/vllm-project/vllm/pull/17045}},
}

@misc{vllm_cve_2025,
  author       = {{GitHub Security Advisory}},
  title        = {{CVE-2025-25183}: {vLLM} Prefix Cache Hash Collision ({GHSA-rm76-4mrf-v9r8})
                  and Timing Side-Channel ({GHSA-4qjh-9fv9-r85r})},
  year         = {2025},
  howpublished = {\url{https://github.com/vllm-project/vllm/security/advisories/}},
}

@misc{openassistant_oasst1,
  title        = {OpenAssistant Conversations Dataset (OASST1)},
  author       = {OpenAssistant},
  year         = {2023},
  howpublished = {\url{https://huggingface.co/datasets/OpenAssistant/oasst1}},
  note         = {Accessed: 2026-05-20}
}

@misc{dataprovenance2024commercialflan_dialog,
  title        = {Commercial FLAN Collection: Dialog},
  author       = {{Data Provenance Initiative}},
  year         = {2024},
  howpublished = {\url{https://huggingface.co/datasets/DataProvenanceInitiative/Commercial-Flan-Collection-Dialog}},
  note         = {Accessed: 2026-05-20}
}

@misc{ultrachat_dataset,
  title        = {UltraChat: An Open-Source Large-Scale Multi-Turn Dialogue Dataset},
  author       = {{OpenBMB}},
  year         = {2023},
  howpublished = {\url{https://huggingface.co/datasets/openbmb/UltraChat}},
  note         = {Accessed: 2026-05-20}
}

@software{ai4privacy_pypi_v050,
  title        = {ai4privacy (Version 0.5.0)},
  author       = {{Ai4Privacy}},
  year         = {2026},
  url          = {https://pypi.org/project/ai4privacy/},
  note         = {PyPI package version 0.5.0, accessed: 2026-05-20}
}

@misc{meta2024llama31,
  title        = {Llama-3.1-70B},
  author       = {{Meta AI}},
  year         = {2024},
  howpublished = {\url{https://huggingface.co/meta-llama/Llama-3.1-70B}},
  note         = {Released July 23, 2024. License: Llama 3.1 Community License}
}

@misc{cve202546570,
  author       = {{National Vulnerability Database}},
  title        = {{CVE-2025-46570}},
  year         = {2025},
  howpublished = {\url{https://nvd.nist.gov/vuln/detail/CVE-2025-46570}},
  note         = {Accessed: 2026-05-29}
}

@InProceedings{pmlr-v267-gu25b,
  title = 	 {Auditing Prompt Caching in Language Model {API}s},
  author =       {Gu, Chenchen and Li, Xiang Lisa and Kuditipudi, Rohith and Liang, Percy and Hashimoto, Tatsunori},
  booktitle = 	 {Proceedings of the 42nd International Conference on Machine Learning},
  pages = 	 {20477--20496},
  year = 	 {2025},
  editor = 	 {Singh, Aarti and Fazel, Maryam and Hsu, Daniel and Lacoste-Julien, Simon and Berkenkamp, Felix and Maharaj, Tegan and Wagstaff, Kiri and Zhu, Jerry},
  volume = 	 {267},
  series = 	 {Proceedings of Machine Learning Research},
  month = 	 {13--19 Jul},
  publisher =    {PMLR},
  url = 	 {https://proceedings.mlr.press/v267/gu25b.html}
}

@software{villegas_pii_extract,
  author = {Paulo Villegas},
  title = {pii-extract-base: Extraction of PII from text chunks},
  year = {2024},
  url = {https://pypi.org/project/pii-extract-base/},
  note = {Python package}
}

@misc{llama32024,
  title        = {The Llama 3.2 Collection of Open Multilingual Large Language Models},
  author       = {{Meta AI}},
  year         = {2024},
  howpublished = {\url{https://huggingface.co/meta-llama/Llama-3.2-1B}},
  note         = {Accessed: 2026-06-02}
}

@misc{ai4privacy2024,
  author       = {Ai4Privacy},
  title        = {{PII}-Masking-200k: A Dataset for Personally Identifiable Information Masking},
  year         = {2024},
  howpublished = {\url{https://huggingface.co/datasets/ai4privacy/pii-masking-200k}},
  note         = {Accessed: 2024}
}

@inproceedings{cheng2025pii,
  title        = {Understanding {PII} Leakage in Large Language Models: A Systematic Survey},
  author       = {Cheng, Shuai and Li, Zhao and Meng, Shu and Ren, Mengxia and Xu, Haitao and Hao, Shuai and Yue, Chuan and Zhang, Fan},
  booktitle    = {Proceedings of the Thirty-Fourth International Joint Conference on Artificial Intelligence (IJCAI-25)},
  pages        = {10409--10417},
  year         = {2025},
  doi          = {10.24963/ijcai.2025/1156}
}

@article{singh2025unmasking,
  title        = {Unmasking the Reality of {PII} Masking Models: Performance Gaps and the Call for Accountability},
  author       = {Singh, Devansh and Narayanan, Sundaraparipurnan},
  journal      = {arXiv preprint arXiv:2504.12308},
  year         = {2025},
  doi          = {10.48550/arXiv.2504.12308}
}

@article{acharya2025comparative,
  title        = {A Comparative Study of Light-weight Language Models for {PII} Masking and their Deployment for Real Conversational Texts},
  author       = {Acharya, Prabigya and Shrestha, Liza},
  journal      = {arXiv preprint arXiv:2512.18608},
  year         = {2025},
  doi          = {10.48550/arXiv.2512.18608}
}

@inproceedings{goguen1982security,
  title        = {Security Policies and Security Models},
  author       = {Goguen, Joseph A. and Meseguer, Jos{\'e}},
  booktitle    = {Proceedings of the IEEE Symposium on Security and Privacy},
  pages        = {11--20},
  year         = {1982},
  organization = {IEEE}
}

@inproceedings{goguen1984unwinding,
  title        = {Unwinding and Inference Control},
  author       = {Goguen, Joseph A. and Meseguer, Jos{\'e}},
  booktitle    = {Proceedings of the IEEE Symposium on Security and Privacy},
  pages        = {75--86},
  year         = {1984},
  organization = {IEEE}
}

@article{denning1976lattice,
  title        = {A Lattice Model of Secure Information Flow},
  author       = {Denning, Dorothy E.},
  journal      = {Communications of the ACM},
  volume       = {19},
  number       = {5},
  pages        = {236--243},
  year         = {1976},
  publisher    = {ACM}
}

@article{volpano1996sound,
  title        = {A Sound Type System for Secure Flow Analysis},
  author       = {Volpano, Dennis and Irvine, Cynthia and Smith, Geoffrey},
  journal      = {Journal of Computer Security},
  volume       = {4},
  number       = {2-3},
  pages        = {167--187},
  year         = {1996}
}

@inproceedings{sabelfeld2003language,
  title        = {Language-Based Information-Flow Security},
  author       = {Sabelfeld, Andrei and Myers, Andrew C.},
  journal      = {IEEE Journal on Selected Areas in Communications},
  volume       = {21},
  number       = {1},
  pages        = {5--19},
  year         = {2003},
  publisher    = {IEEE}
}
\end{document}